\documentclass[pdftex,5p,preprint,final,times,sort&compress]{elsarticle}
\usepackage{amsmath}
\usepackage{amssymb}
\usepackage{xspace}
\usepackage{hyperref}
\newcommand{\arcdeg}{\ifmmode^\circ\else$^\circ$\fi}
\newcommand{\arcmin}{\ifmmode'\else$'$\fi}
\newcommand{\arcsec}{\ifmmode''\else$''$\fi}
\newcommand{\plotone}{\centering\includegraphics[width=.9\linewidth]}
\newcommand{\phn}{\ifmmode\phantom{0}\else$\phantom{0}$\fi}
\newcommand{\phm}{\ifmmode\phantom{-}\else$\phantom{-}$\fi}
\newcommand{\tablenotemark}[1]{$^\mathrm{#1}$}
\newcommand{\tablenotetext}[2]{\parbox{\linewidth}{$^\mathrm{#1}$\,\footnotesize #2}\par}
\newcommand\tomoe{Tomo-e Gozen\xspace}
\newcommand\tomoepm{Tomo-e PM\xspace}
\usepackage{color}
\definecolor{20180710}{rgb}{0,0,0}
\newcommand{\revise}[2]{\textcolor{#1}{#2}}
\begin{document}
\title{Luminosity Function of Faint Sporadic Meteors measured with a Wide-Field CMOS mosaic camera \tomoepm}
\author[ioa,kiso]{Ryou Ohsawa}
\ead{ohsawa@ioa.s.u-tokyo.ac.jp}
\author[ioa,jst]{Shigeyuki Sako}
\author[lih]{Yuki Sarugaku}
\author[cps]{Fumihiko Usui}
\author[jaxa]{Takafumi Ootsubo}
\author[sokendai]{Yasunori Fujiwara}
\author[nms]{Mikiya Sato}
\author[naoj]{Toshihiro Kasuga}
\author[naoj]{Ko Arimatsu}
\author[naoj]{Jun-ichi Watanabe}
\author[ioa]{Mamoru Doi}
\author[ioa,kiso]{Naoto Kobayashi}
\author[ioa]{Hidenori Takahashi}
\author[ioa]{Kentaro Motohara}
\author[ioa]{Tomoki Morokuma}
\author[ioa]{Masahiro Konishi}
\author[kiso]{Tsutomu Aoki}
\author[kiso]{Takao Soyano}
\author[kiso]{Ken'ichi Tarusawa}
\author[kiso]{Yuki Mori}
\author[ioa,kiso]{Yoshikazu Nakada}
\author[ioa]{Makoto Ichiki}
\author[ioa]{Noriaki Arima}
\author[ioa]{Yuto Kojima}
\author[ioa]{Masahiro Morita}
\author[resceu]{Toshikazu Shigeyama}
\author[tohoku]{Yoshifusa Ita}
\author[tohoku]{Mitsuru Kokubo}
\author[tohoku]{Kazuma Mitsuda}
\author[kyoto]{Hiroyuki Maehara}
\author[konan]{Nozomu Tominaga}
\author[naoj]{Takuya Yamashita}
\author[ism]{Shiro Ikeda}
\author[ism]{Mikio Morii}
\author[bsgc]{Seitaro Urakawa}
\author[bsgc]{Shin-ichiro Okumura}
\author[jaxa]{Makoto Yoshikawa}
\address[ioa]{Institute of Astronomy, Graduate School of Science, University of Tokyo, 2-21-1 Osawa, Mitaka, Tokyo 181-0015, Japan}
\address[kiso]{Kiso Observatory, Institute of Astronomy, Graduate School of Science, The University of Tokyo, 10762-30 Mitake, Kiso-machi, Kiso-gun, Nagano 397-0101, Japan}
\address[lih]{Koyama Astronomical Observatory, Kyoto Sangyo University, Kamigamo Motoyama, Kita, Kyoto, Kyoto Prefecture 603-8555, Japan}
\address[cps]{Center for Planetary Science, Graduate School of Science, Kobe University, 7-1-48, Minatojima-Minamimachi, Chuo-ku, Kobe 650-0047, Japan}
\address[jaxa]{Department of Space Astronomy and Astrophysics, Institute of Space and Astronautical Science (ISAS), Japan Aerospace Exploration Agency (JAXA), 3-1-1 Yoshinodai, Chuo-ku, Sagamihara, Kanagawa 252-5210, Japan}
\address[nms]{The Nippon Meteor Society, Japan}
\address[sokendai]{SOKENDAI (The Graduate University for Advanced Studies), 10-3 Midoricho, Tachikawa, 190-8518 Tokyo, Japan}
\address[naoj]{National Astronomical Observatory of Japan, 2-21-1 Osawa, Mitaka, Tokyo 181-8588, Japan}
\address[resceu]{Research Center for the Early Universe, School of Science Bldg. No.4, University of Tokyo, Hongo, 7-3-1, Bunkyo-ku, Tokyo, 113-0033, Japan}
\address[tohoku]{Astronomical Institute, Tohoku University, 2 Chome-1-1 Katahira, Aoba Ward, Sendai, Miyagi Prefecture 980-8577, Japan}
\address[konan]{Department of Physics, Faculty of Science and Engineering, Konan University, 8 Chome-9-1 Okamoto, Higashinada Ward, Kobe, Hyogo Prefecture 658-0072, Japan}
\address[kyoto]{Astronomical Observatory, Graduate School of Science, Kyoto University, Yoshida-honmachi, Sakyo-ku, 606-8501 Kyoto, Japan}
\address[ism]{The Institute of Statistical Mathematics, 10-3 Midori-cho, Tachikawa, Tokyo 190-8562, Japan}
\address[bsgc]{Japan Spaceguard Association, Bisei Spaceguard Center, 1716-3 Ookura, Bisei-cho, Ibara-shi, Okayama, 714-1411, Japan}
\address[jst]{Precursory Research for Embryonic Science and Technology (PRESTO), Japan Science and Technology Agency (JST), 2-21-1 Osawa, Mitaka, Tokyo, 181-0015, Japan}
\begin{abstract}
Imaging observations of faint meteors were carried out on April 11 and 14, 2016 with a wide-field CMOS mosaic camera, \tomoepm, mounted on the 105-cm Schmidt telescope at Kiso Observatory, the University of Tokyo. \tomoepm, which is a prototype model of \tomoe, can monitor a sky of \revise{20180710}{${\sim}1.98\,\mathrm{deg^2}$} at 2\,Hz. The numbers of detected meteors are 1514 and 706 on April 11 and 14, respectively. The detected meteors are attributed to sporadic meteors. Their absolute magnitudes range from $+4$ to $+10\,\mathrm{mag}$ in the $V$-band, corresponding to about $8.3{\times}10^{-2}$ to $3.3{\times}10^{-4}\,\mathrm{g}$ in mass. \revise{20180710}{The present magnitude distributions we obtained are well explained by a single power-law luminosity function with a slope parameter $r = 3.1{\pm}0.4$ and a meteor rate $\log_{10}N_0 = -5.5{\pm}0.5$. The results demonstrate a high performance of telescopic observations with a wide-field video camera to constrain the luminosity function of faint meteors. The performance of \tomoe is about two times higher than that of \tomoepm. A survey with \tomoe will provide a more robust measurement of the luminosity function.}
\end{abstract}
\maketitle
\section{Introduction}
\label{sec:intro}
The solar system is composed of a variety of objects with different sizes: the Sun, the planets and their moons, other small bodies including dwarf planets, asteroids and comets, and \revise{20180710}{dust grains in interplanetary space}. The size of the small bodies in the solar system covers a wide dynamic range from 1,000\,km (\textit{e.g.}, Ceres) down to about 10\,nm ($\beta$-asteroids). The size distribution of the small bodies provides important information to understand the origin and the evolution of the solar system.
The amount of the interplanetary materials incoming the Earth is about $10^5\,\mathrm{kg\,day^{-1}}$. The particles in the mass range of $10^{-9}$--$10^{-2}\,$g are the major contributors of the incoming mass flux \citep{grun_collisional_1985}. Properties of such small particles are essential to understand what kind of materials fall into the Earth.
Particles smaller than about $10^{-6}\,$g \revise{20180710}{have been detected in place with dust counters on satellites and spacecrafts} \citep{gruen_ulysses_1992,grun_galileo_1992,gurnett_micron-sized_1997,szalay_student_2013}, while the number density of the large particles is so small that their size distribution is difficult to measure by in-situ observations.
The small particles incoming the Earth are observed as meteors, interacting with the Earth's atmosphere and \revise{20180710}{convert part of their kinetic energy into emission lines} \citep{baldwin_ablation_1971,ceplecha_meteor_1998,hawkes_quantitative_1975,jones_effects_1966,popova_meteoroid_2004}. Thus, the mass of the incoming particle can be derived from measuring the brightness of the meteor with an assumption of an energy conversion efficiency. The Earth works as an extremely wide-area \revise{20180710}{dust} detector, providing an indirect measurement of the size distribution of \revise{20180710}{interplanetary dust} around the Earth.
Meteors in the range between $-5$ and $+4\,\mathrm{mag}$. have been detected by large meteor survey networks, for example, Cameras for Allsky Meteor Surveillance (CAMS), and SonotaCo network \citep{jenniskens_meteor_2017}. CAMS obtained more than 135,000 meteor orbits in 2017. In this magnitude range, about 36\% of meteors belong to some meteor showers \citep{jenniskens_cams_2016-2}. The luminosity function of meteors in this magnitude range was intensively investigated by \citet{hawkins_influx_1958} based on a large survey with Super-Schmidt cameras. They showed that the meteor luminosity function was well-approximated by a power-law function. The meteors detected by these surveys, however, are mostly caused by particles of larger than ${\sim}1\,\mathrm{g}$. Observations of meteors fainter than $+5\,\mathrm{mag}$ are required to investigate the size distribution of \revise{20180710}{interplanetary dust}.
\citet{cook_flux_1980} obtained more than 2,000 sporadic meteors in four nights using a 10-m optical reflector and a photo multiplier. The photographic magnitudes of the detected meteors ranged between $+7$ and $+12\,\mathrm{mag}$. Comparing with literature, they concluded that the meteor luminosity function was well approximated by a power-law function between $-2.4$ to $+12.0\,\mathrm{mag}$. They observed meteors as light pulses since they used the photo multiplier instead of a camera. Thus, the information on the meteor trajectories was not obtained and the atmospheric correction to compensate a possibly unstable weather condition was not applied in real time. \citet{clifton_television_1973} and \citet{hawkes_television_1975} observed faint meteors down to about $8$ and $6\,\mathrm{mag}$, respectively, in television observations. Imaging observations of faint meteors using a wide-field video camera are favored to examine the meteor luminosity function in detail.
Radar observations detect meteors caused by particles in the mass range of about $10^{-6}$--$10^{-2}\,$g. The Canadian Meteor Orbit Radar has detected more than 3,000,000 meteors down to $10^{-4}$\,g in seven years \citep{brown_meteoroid_2008,brown_meteoroid_2010}. Large aperture radars can observe meteor head echoes. \revise{20180710}{Interplanetary dust grains} of about $10^{-6}\,g$ can be detected by these radars. \citet{kero_2009-2010_2012} obtained more than 100,000 meteors using the Shigaraki Middle and Upper atmosphere radar in Japan. Both the trajectories and the radar cross-sections of meteors were obtained simultaneously. This is a great advantage of radar head echo observations. On the other hand, the radar cross-sections are not easy to convert to the brightness of the meteors, or the mass of the meteoroids. Still, there is a need for optical observations of faint meteors.
The \revise{20180710}{interplanetary dust grains} in the mass range of $10^{-6}$--$10^{-3}\,\mathrm{g}$ correspond to about $+7$--$14\,\mathrm{mag}$ in optical observation. To obtain images of such faint meteors, we need a large photon-collecting mirror, a wide-field optics, and a high-sensitive video camera \citep{kresakova_activity_1955,poleski_1996-2001_2008}. \revise{20180710}{There are only a handful of studies on meteors detected with large telescopes. \citet{pawlowski_leonid_1998} detected 151 Leonid meteors down to about $13\,\mathrm{mag}$ with the Liquid Mirror Telescope in Cloudcroft Observatory, to constrain the mass distribution index of the Leonid shower. \citet{iye_suprimecam_2007} serendipitously detected sporadic meteors with \textit{Subaru}/SuprimeCam and constrained the diameter of the collisionally excited tubes caused by the meteors. \citet{bektesevic_linear_2017} developed a new method to discover faint meteors obtained in large optical surveys, such as the Sloan Digital Sky Survey. Currently, no facility conducts regular observations of faint meteors with a telescope larger than $1\,\mathrm{m}$.}
We carried out observations of faint meteors with a wide-field mosaic CMOS camera \tomoepm in 2016. This paper summarizes the results of the observations. This paper is organized as follows: details of observations are summarized in Seciton~\ref{sec:observation}; data reduction and observed meteors are presented in Section~\ref{sec:result}; the luminosity function of faint meteors is discussed in Section~\ref{sec:discussion}; we summarize the paper in Section~\ref{sec:conclusion}.
\section{Observations}
\label{sec:observation}
\subsection{\tomoepm}
\label{sec:tomoe-pm}
\tomoe is a wide-field camera being developed in Kiso observatory, the University of Tokyo, which will be the world largest video camera for astronomy. \tomoe, mounted on the 105-cm Kiso Schmidt Telescope\footnote{Kiso Observatory is located in $35\arcdeg 47\arcmin 38.7\arcsec$\,N and $137\arcdeg 37\arcmin 42.2\arcsec$\,E, and at the altitude of $1130\,\mathrm{m}$.} in Kiso observatory, will continuously monitor a \revise{20180710}{$20\,\mathrm{deg^2}$} area at 2\,Hz with 84 CMOS sensors developed by Canon Inc. The designed pixel scale is about \revise{20180710}{$1.19{\arcsec}{\times}1.19{\arcsec}$}. The limiting magnitude is estimated to be about $18.5\,\mathrm{mag}$. in the $V$-band in the 2\,Hz observation. The limiting magnitude for meteors will be about $13\,\mathrm{mag}$. in the $V$-band, on the assumption that meteors moves at \revise{20180710}{$10\arcdeg \mathrm{s}^{-1}$, which corresponds to the angular velocity of a meteor at zenith, at 100\,km above an observer, at a velocity of $36\,\mathrm{km\,s^{-1}}$, and at an incident angle of $30\deg$}.
As a pilot project, we developed a prototype of \tomoe (\textit{hereafter}, \tomoepm), equipped with the 8 CMOS sensors \citep{sako_development_2016}. \tomoepm is mounted on the 105\,cm Schmidt telescope at Kiso Observatory, the University of Tokyo. The pixel scale is about \revise{20180710}{$1.19{\arcsec}{\times}1.19{\arcsec}$}. The detectors are aligned in a line along the direction of the right ascension with some gaps (\textit{see}, \citet{sako_development_2016}). The total area of the field-of-view is about \revise{20180710}{$1.98\,\mathrm{deg^2}$}. All the detectors are synchronously operated by the same control signal. The maximum frame rate is 2\,Hz. The overhead time due to readout is almost negligible thanks to rolling shutter. A NTP synchronized time is recorded as a time-stamp in a FITS header, although the time-stamp is not synchronized with a shutter timing. The time-stamp is as accurate as ${\sim}1\,\mathrm{s}$. The size of the imaging area is $2000{\times}1128$ pixels. The experimental observations were successfully completed, to confirm that \tomoepm achieved the designed sensitivity.
\subsection{Observations}
\label{sec:observations}
Meteor observations with \tomoepm were carried out on April 11 and 14, 2016 (UT). Details of the observations are summarized in Table~\ref{tab:obslog}. The sky was dark and clear on April 11, while part of the observations were \revise{20180710}{affected by} clouds on April 14.
Meteors are detected as a streak in the 2\,Hz observation, while artificial satellites and space debris are also detected as streaks. To eliminate these contaminations, the observations were conducted in UT 12--18, ${\pm}3\,$hours around midnight in JST, and the observed regions were set within the Earth's shadow. Since the shadow moves during a night, the observed regions were relocated in every two hours.
The captured movie data were compiled into FITS data cubes in every 3 minutes (\textit{hereafter}, an observation unit). Thus, each FITS data cube was composed of 360 frames. In total, 808 observation units (290,880 frames) were obtained on April 11, while 880 observation units (316,800 frames) were obtained on April 14.
\begin{table*}[tp]
\centering
\caption{Observation Logs}
\label{tab:obslog}
\begin{tabular}{lrr}
\hline\hline
Date
& 2016-04-11 & 2016-04-14 \\
Weather Condition
& Clear Sky & Partly Clouded \\
Lunar Age
& 3.6 & 6.6 \\
Elevation Range
& $13.7\arcdeg$--$37.9\arcdeg$ & $10.1\arcdeg$--$37.1\arcdeg$ \\
Typical Zero Magnitude
& $25.5\,\mathrm{mag}$ & $25.0\,\mathrm{mag}$ \\
Total Observing Time
& ${\sim}5.1\,$hours & ${\sim}5.5\,$hours \\
Number of Exposures
& 290,880 & 316,800 \\
Number of Detected Meteors
& 1,514 & 706 \\
Apparent Magnitudes\tablenotemark{\dagger}
& $4.5$--$12.5\,\mathrm{mag}$ & $5.5$--$11.0\,\mathrm{mag}$ \\
\hline
\end{tabular}
\tablenotetext{\dagger}{Video rate magnitudes defined in \citet{iye_suprimecam_2007}.}
\end{table*}
\section{Results}
\label{sec:result}
\subsection{Data Reduction}
\label{sec:reduction}
\begin{figure*}[p]
\centering
\plotone{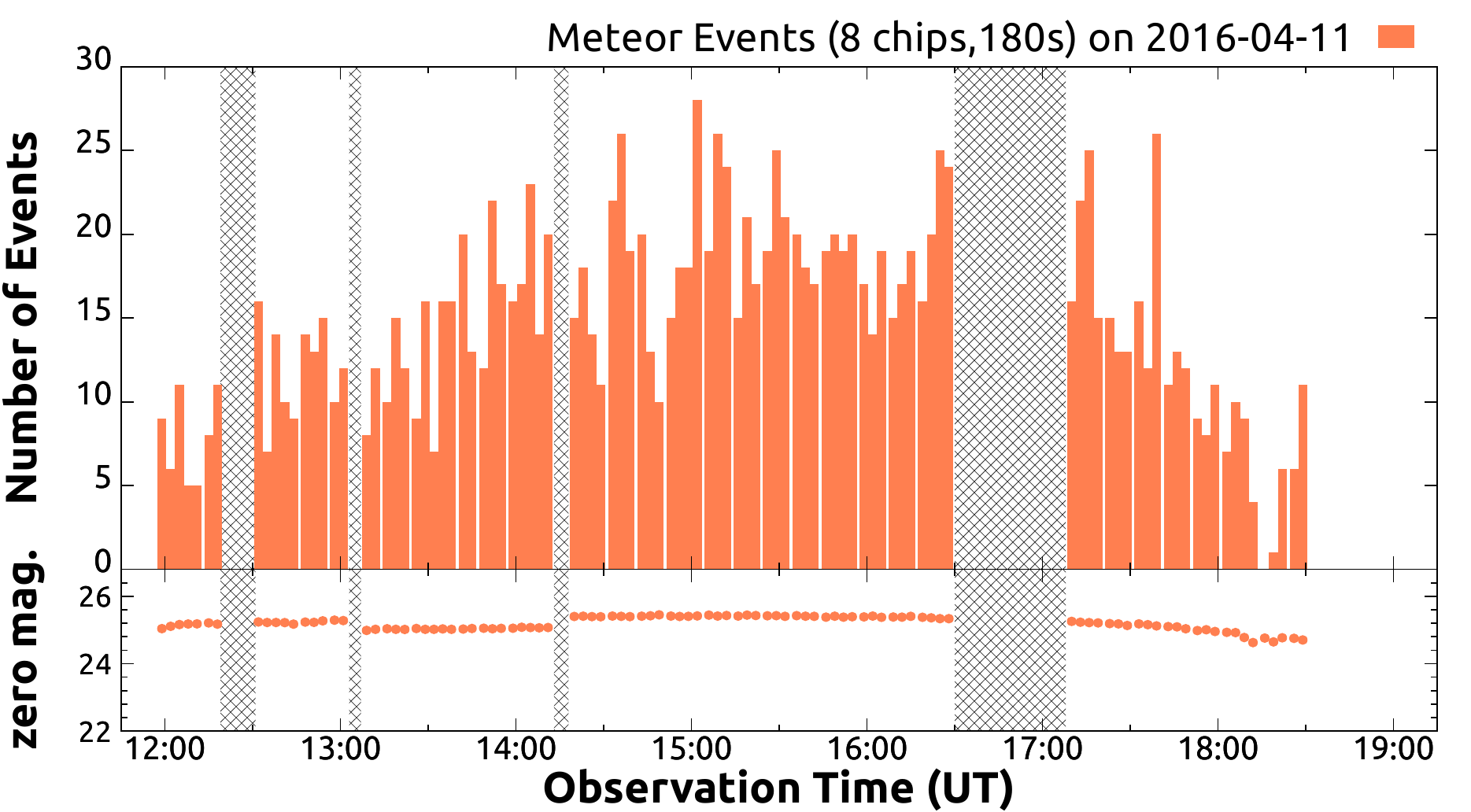}
\plotone{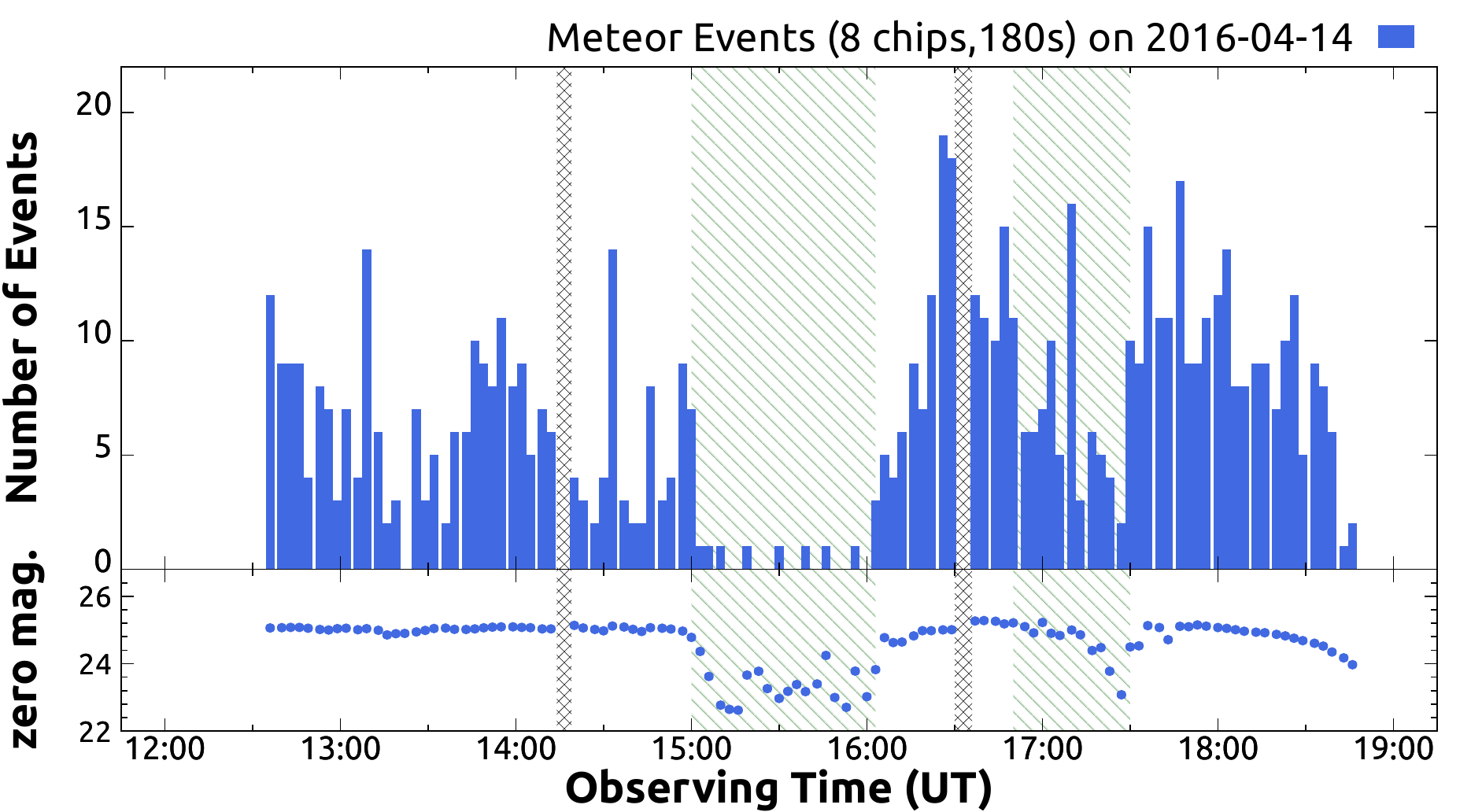}
\caption{Time variation in the number of meteors and the magnitude zero-point. The periods of cloudy weather are indicated by the green hatched regions. The gray cross-hatched regions mean no data due to pointings or troubles.}
\label{fig:evrates}
\end{figure*}

Streak-like signals were extracted from the movie data cube with the slant-and-compress algorithm developed by \citet{ohsawa_development_2016}. First, background emission was subtracted and stellar signals were masked. A frame was slanted by a certain angle and compressed into a one-dimensional array with the normalization such that the noise level should be uniform. A set of one-dimensional arrays were stacked into a two-dimensional array. The streak-like signals in the original image were converted as compact sources in the converted array. The signal-to-noise ratio of faint meteors in the original image was enhanced by about $\sqrt{n}$, where $n$ was the number of pixels along the streak, by integrating signals along with the streak. Refer to \citet{ohsawa_development_2016} for details of the detection algorithm.
A number of meteor candidates were detected by the algorithm. Then, genuine meteor events were counted by humans. The numbers of the real meteor events were 2002 and 926 on April 11 and 14, respectively. Some meteors were doubly counted due to the rolling shutter or passing through multiple detectors. Those duplicated events were merged. Finally, 1514 and 706 events were extracted as the individual meteor events detected on April 11 and 14, respectively. \revise{20180710}{Although no prominent meteor shower activity was predicted at that time, there could be some contributions from April Lyrids and $\eta$-Aquariids. The separation angles from the radiant points of these showers were calculated. The fractions of the meteors which might pass through within 3\arcdeg of the radiants were at most 6\% and 3\% for April Lyrids and $\eta$-Aquariids, respectively. These fractions were consistent with the number of sporadic meteors randomly coming from the radiants. We concluded that the contribution from meteor showers was negligible and the detected meteors were attributed to sporadic meteors.}
The time variations in the number of the meteors on April 11 and 14 are illustrated in the top and bottom panels of Figure~\ref{fig:evrates}, respectively. The bar charts show the numbers of the meteors in the observation unit. The circles below the bar charts illustrates the time variations in the magnitude zero-point averaged over the eight detectors, indicating weather conditions in the observation unit. The small number of the detections on April 14 was simply attributed to the poor weather conditions.
The brightness of the meteors was measured as follows. The projected profile of the meteor was obtained by averaging the signal along with the streak. The projected profile was broad and sometimes double-peaked since meteors were usually defocused \citep{bektesevic_linear_2018}. The line-intensity of the meteor, $\tilde{I}$ in units of $\mathrm{ADU\,pix^{-1}}$, was measured by fitting the projected profile with a combination of Gaussian profiles. Since the angular velocity of the meteors were not measured by \tomoepm observations, the line-intensity $\tilde{I}$ was not directly compared with the intensities of nearby stars. Thus, we adopted the video-rate magnitude defined in \citet{iye_suprimecam_2007}. Assuming that all the meteors moved at \revise{20180710}{an angular velocity of $10\arcdeg s^{-1}$}, the travel distance of the meteors in $0.5\,$s was estimated to be $1.5{\times}10^3$ pixels, where $\theta$ was the pixel scale of \tomoepm in units of arcsecond. The total amount of the signals emitted by the meteor in $0.5\,$s was obtained by ${I} = 1.5{\times}10^3\tilde{I}\,\mathrm{ADU}$, which was transformed into the $V$-band magnitude by comparing with the intensities of nearby stars. The ranges of the estimated magnitudes are listed in Table~\ref{tab:obslog}. The meteors as faint as about $12.5\,\mathrm{mag}$ in the $V$-band were detected on April 11. Although the weather condition on April 14 was poorer than that on April 11, the meteors as faint as about $11.0\,\mathrm{mag}$ were successfully detected.
\subsection{System Efficiency}
\label{sec:syseff}
\begin{figure*}
\centering
\plotone{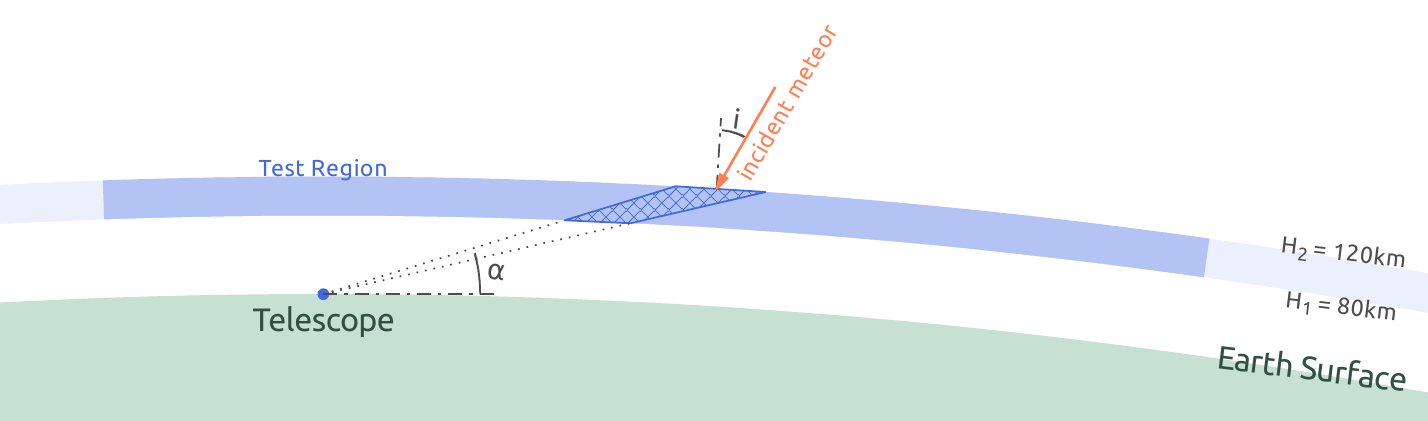}
\caption{Definition of the meteor detectable volume. The blue dot is the location of a telescope, which is observing a sky at the elevation $\alpha$. The thin blue region indicates the meteor detectable layer. The blue cross-hatched region indicates the meteor detectable volume. The thick blue region is the test region, where meteors are randomly generated. A generated meteor with the incident angle $\theta$ is indicated by the red arrow.}
\label{fig:schematic-collect}
\end{figure*}
The meteor-collecting power of a observing system is usually evaluated in units of area. \citet{koschack_determination_1990} evaluated the effective meteor-collecting power from the area of the field-of-view projected onto a meteor level at an altitude of $100\,\mathrm{km}$ weighted by distance. In observation with naked eyes (visual observation), a field-of-view was as large as about $52.5\arcdeg$ \citep{koschack_determination_1990}. The estimated effective meteor-collecting area (\textit{hereafter}, EMCA) in visual observation is $24,400\,\mathrm{km^2}$ when observing at zenith, whereas that area becomes $20,770\,\mathrm{km^2}$ when observing with an elevation angle of $50\arcdeg$, in the case that a meteor population index\footnote{The number of meteors brighter than a magnitude $m$ in visible wavelengths should follow $N({<}m) \propto r^m$.} $r$ is 3.5. In Koschack's calculation, the region where meteors are observable is approximated by a thin spherical shell. Hereafter, Koschack's formalism is referred as to the thin shell approximation.
In meteor observations with telescopes, the field-of-view is usually smaller than 10\arcdeg. In such cases, the thin shell approximation is not valid. \citet{kresakova_activity_1955} evaluated the EMCA by statistically estimating the averaged angular length of meteors. This method requires a large number of meteors. Since the efficiency in meteor detection with \tomoepm changes with elevation and \tomoepm has the multiple image sensors, their method is not applicable to our observations. Thus, we evaluated the EMCA of \tomoepm by a Monte Carlo simulation.
Figure~\ref{fig:schematic-collect} schematically describes the configuration of calculation. When \revise{20180710}{dust grains} penetrate into the upper atmosphere, they are heated by interaction and become bright in optical wavelengths at a certain altitude, observed as meteors. Then, when the \revise{20180710}{grain} penetrates into a lower atmospheric layer, they cease to be bright. Here, we define an upper and lower limit of an altitude where meteors are observable in optical wavelengths by $H_2$ and $H_1$, respectively. Thus, the region where meteors are observable in optical wavelengths is approximated by a spherical shell with a depth of $H_2{-}H_1$ (\textit{hereafter}, meteor shell). In Figure~\ref{fig:schematic-collect}, the meteor shell is indicated by the blue shaded region.
The region captured by a camera with a single rectangular field-of-view is defined by
\begin{equation}
\label{eq:mov}
\vec{x} \in \left\{~
c_0\vec{n}_0+c_1\vec{n}_1+c_2\vec{n}_2+c_3\vec{n}_3
~\middle|~
c_m > 0 ~\text{for}~ m=0,1,2,3
~\right\},
\end{equation}
where $\vec{n}_m (m=0,1,2,3)$ is a unit vector to point a corner of the field-of-view. The intersection of the meteor shell and the field-of-view is defined as a meteor-collecting volume (MCV). The MCV is approximated by a truncated pyramid as shown by the cross-hatched region in Figure~\ref{fig:schematic-collect}. This approximation does not affect results significantly as long as the dimension of the field-of-view is smaller than about $1\arcdeg$. The distance to the MCV is represented by the distance to its geometric center. For a system with multiple image sensors such as \tomoepm, the union of the MCVs for each sensor is defined as the MCV of the system.
An incident meteor is defined by two properties: an incident point $\vec{p}$ and a pair of incident angles $(\omega,i)$. The incident point $\vec{p}$ means the location of \revise{20180710}{a dust grain} when they enter the meteor shell. The incident angles correspond to the velocity of \revise{20180710}{a grain} relative to the Earth. The angle $\omega$ is an azimuthal angle, while the angle $i$ is an incident angle measured from the normal vector of the meteor shell at $\vec{p}$. Incident meteors are illustrated by the orange arrows in Figure~\ref{fig:schematic-collect}. The evolution of the velocity by interaction with atmosphere is not taken into account. Thus, we assume that the direction of a meteor's velocity is constant.
Incident meteors are randomly generated on the test region, which is a sufficiently large part of the meteor shell, with a surface density of $\Sigma$ at an altitude of $H_2$. An effective meteor-collecting area (EMCA) at a nominal altitude of $100\,\mathrm{km}$ is defined by
\begin{equation}
\label{eq:mca}
A = \frac{N}{\Sigma}\left(\frac{100\,\mathrm{km}}{H_2}\right)^2
= N\frac{A_0}{N_0}\left(\frac{100\,\mathrm{km}}{H_2}\right)^2,
\end{equation}
where $N$ is the number of the meteors which enter the MCV, $A_0$ is the total area where the meteors are generated, $N_0$ is the total number of the generated meteor. We developed a code to calculate an EMCA for a given observing configuration. The shape, the elevation angle $\theta$, and the position angle $\phi$ of pointing are given. Incident meteors are generated until $N$ exceeds a threshold $\hat{N}$. Since the detections are given by a Poisson process, the statistical error of $A$ becomes negligible for sufficiently large $N_0$.
When the field-of-view is as large as ${\sim}50\arcdeg$ in diameter, the depth of the meteor shell becomes negligible. Thus the EMCA calculated here becomes asymptotically identical to that in the thin shell approximation.
As the simplest case, we assumed that the number density of \revise{20180710}{interplanetary dust} around the Earth was uniform and the velocity of the \revise{20180710}{interplanetary dust grains} relative to the Earth was isotropic. Thus, the distribution of $\vec{p}$ was uniform on the top surface of the meteor shell, the distribution of $\omega$ distributed uniformly in $[0,\,2\pi)$, and the distribution of $i$ was set to be proportional to $\cos{i}$. In calculation, we adopted $H_1 = 80\,\mathrm{km}$ and $H_2 = 120\,\mathrm{km}$. The EMCA does not significantly depend on the choice of $H_1$ and $H_2$ (see, \ref{sec:depth_meteor_shell}). The threshold $\hat{N}$ was set to $100,000$ so that the statistical errors become negligible. Note that the effective meteor-collecting area calculated based on our assumption may contain a systematic error since the radiant distribution of sporadic meteor is not uniform \citep{jones_radiant_1994}, and may be incorrect especially for meteor showers.
\begin{figure*}
\centering
\plotone{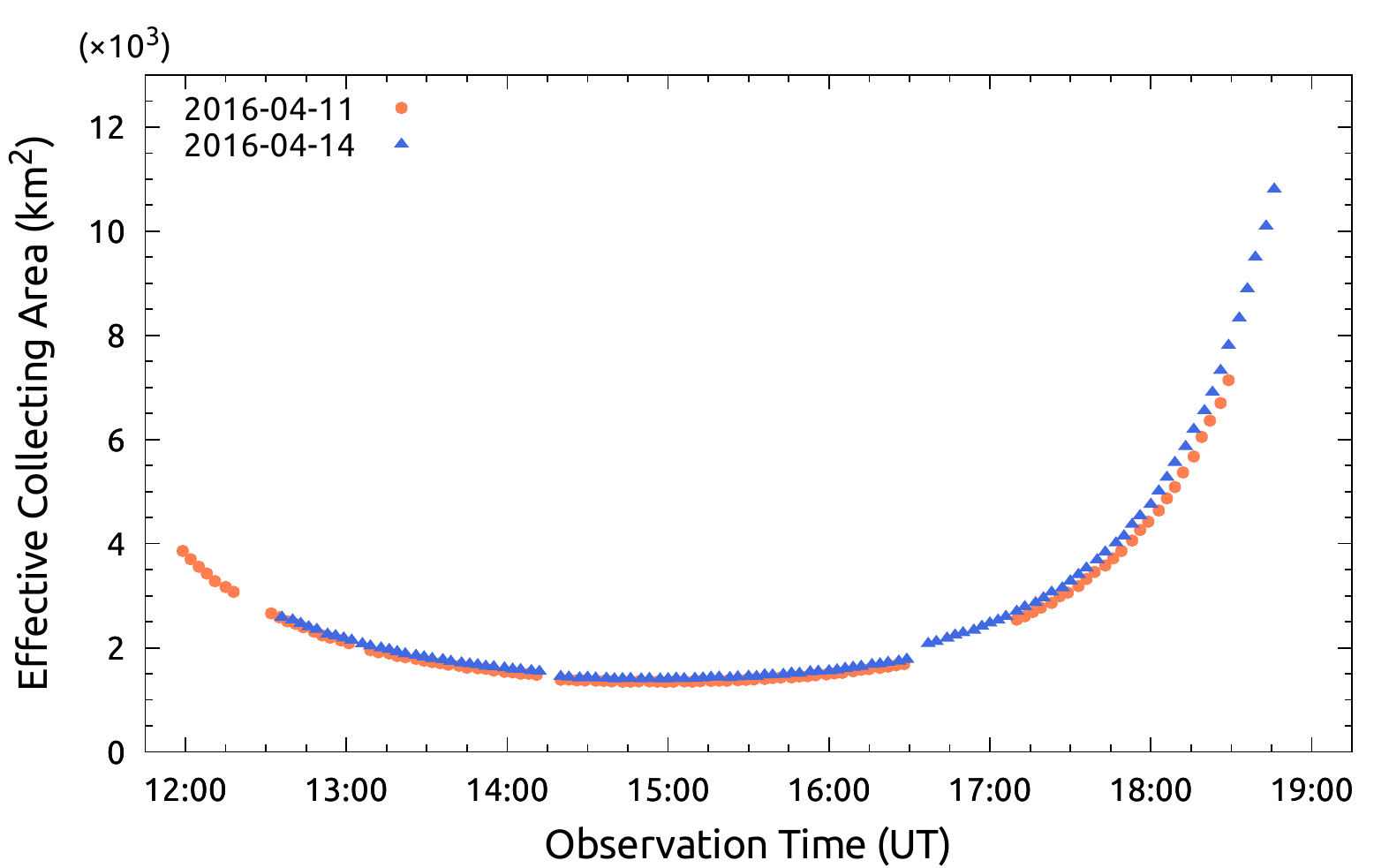}
\caption{Time variation in the meteor-collecting areas on April 11 (the red circles) and April 14 (the blue triangles).}
\label{fig:collecting-area}
\end{figure*}
The EMCAs of \tomoepm estimated with a Monte-Carlo simulation are shown in Figure~\ref{fig:collecting-area}. Random errors are as small as the symbols. Figure~\ref{fig:collecting-area} indicates that the meteor-collecting area of \tomoepm changed from about $1,000\,\mathrm{km}$ to $10,000\,\mathrm{km}$ with elevation. \revise{20180710}{Note that the EMCAs calculated here do not include the effect of the distance between the observer and meteors. When the elevation angle is low, a typical distance to meteors becomes large and faint meteors are not detected. Thus, the number of the detected meteors tends to decrease with decreasing elevation angle. Figure~\ref{fig:evrates} shows that the number of the detected meteors peaked around 15:00 UT on April 11. Detailed discussion is presented in \ref{sec:rmca}. The meteor travel distances and a possible bias in the meteor arrival directions are discussed in \ref{sec:travel_distances} and \ref{sec:position_angles}, respectively. }
\section{Discussion}
\label{sec:discussion}
\subsection{Absolute Magnitude Distribution}
\label{sec:absmags}
\begin{figure*}[tp]
\centering
\plotone{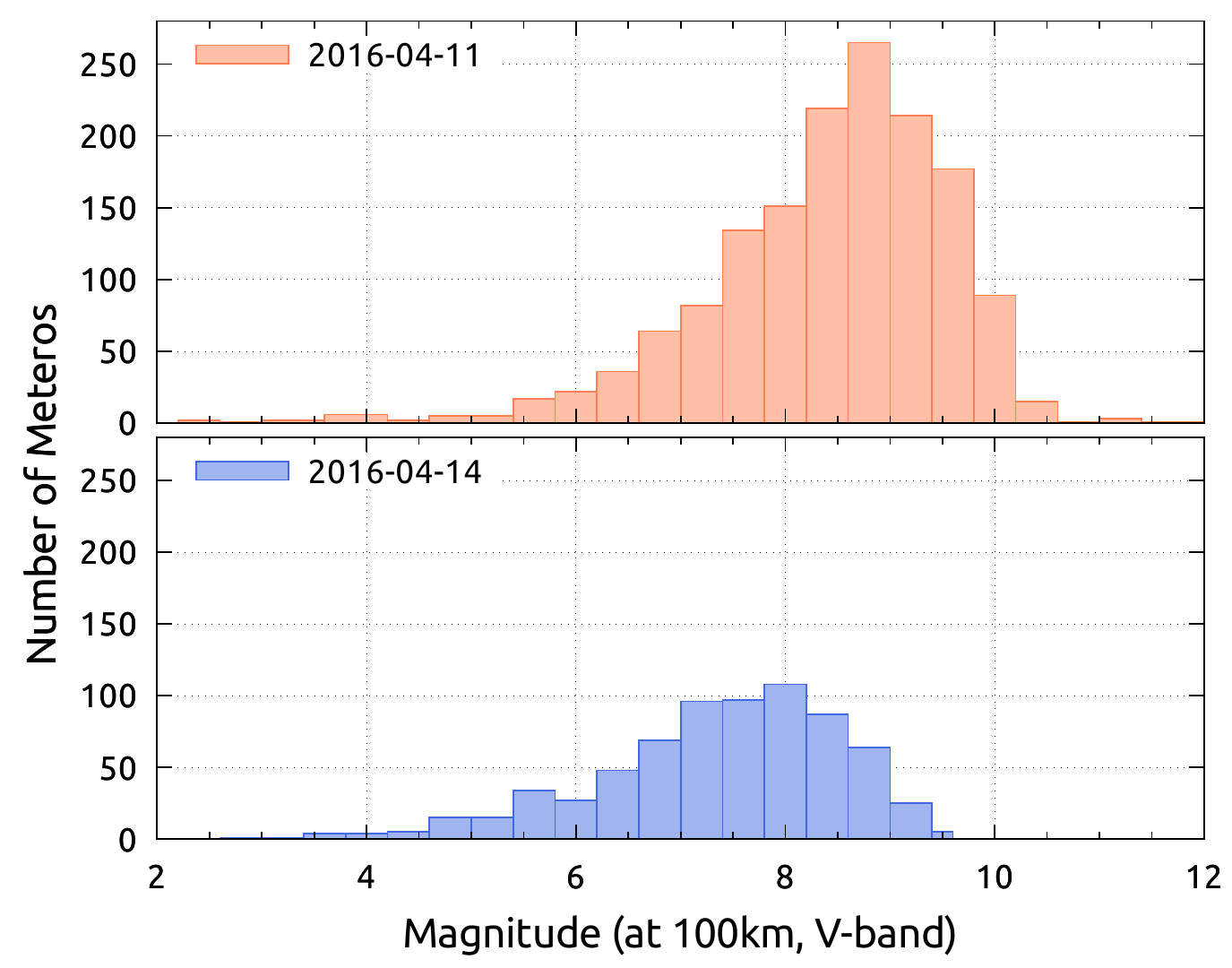}
\caption{Distributions of the absolute magnitudes of the meteors detected by \tomoepm. The top and bottom panels respectively shows the distributions on April 11 and 14, 2016.}
\label{fig:hist}
\end{figure*}
The observed video-rate magnitudes $m_{V}$ were converted into absolute magnitudes $M_{V}$, which correspond to $V$-band magnitudes at the distance of $100\,\mathrm{km}$, by assuming that the distances to meteors were identical to those to the meteor detectable volume:
\begin{equation}
\label{eq:mag100km}
M_V = m_V
+ 5.0\log_{10}\frac{D(\alpha)}{100\,\mathrm{km}}.
\end{equation}
Figure~\ref{fig:hist} shows the distributions of the absolute magnitudes. Meteros of $M_V \lesssim 10$ and $9\,\mathrm{mag}$ were detected on April 11 and 14, respectively. The histograms increase exponentially with increasing magnitudes, and then start to decrease at magnitudes of 8th or 9th. The decrease is basically attributed to sensitivity limits. Detailed discussion is given in Section~\ref{sec:lf_analysis}.
\citet{jacchia_physical_1955} derived an equation to convert the magnitude of a meteor to the mass of a meteoroid, which were calibrated based on observations with Super-Schmidt cameras. A modified equation was given by \citet{jenniskens_meteor_2006}, which derives the mass of a meteoroid from the $V$-band magnitude of a meteors:
\begin{equation}
\label{eq:jenniskens}
\log_{10}{M_g} = 6.31 - 0.40M_V
- 3.92\log_{10}V_\infty - 0.41\log_{10}\sin(h_r),
\end{equation}
where $M_g$ is the mass of the meteoroid in units of grams, $M_V$ is the $V$-band magnitude of the meteor, $V_\infty$ is the incident velocity of the meteoroid in units of $\mathrm{km\,s^{-1}}$, and $h_r$ is the elevation of the radiant point of the meteor. By assuming that $V_\infty$ is $30\,\mathrm{km\,s^{-1}}$ and $h_r$ is $90\arcdeg$, the mass of the meteors we detected ranges from about $8.3{\times}10^{-2}\,\mathrm{g}$ ($4\,\mathrm{mag}$) to about $3.3{\times}10^{-4}\,\mathrm{g}$ ($10\,\mathrm{mag}$).
\subsection{Luminosity Function Analysis}
\label{sec:lf_analysis}
Generally, a luminosity function of visible meteors is well approximated by an exponential function \citep{hawkins_influx_1958}:
\begin{equation}
\label{eq:lf}
\log_{10} N({<}M) = \log_{10}N_0 + M\log_{10}r,
\end{equation}
where $N({<}M)$ and $N_0$ are the event rates of meteors brighter than $M$-th and zero-th magnitude, respectively.
The present results are a composite of the observations in different conditions in terms of the meteor-collecting area and the limiting magnitudes. Thus, a naive fitting of the apparent magnitude distribution to derive luminosity functions may bring biased results. We introduce a statistical model to robustly estimate the slope parameter $r$ and the meteor rate $\log_{10}N_0$.
We assume that the luminosity function of meteors exactly follows Equation~(\ref{eq:lf}) and observational conditions are constant within the observation unit. The number of detectable meteors per observation unit is expected by
\begin{equation}
\label{eq:expnum}
\left\{
\begin{array}{ll}
\log_{10}\tilde{n}^i &= \log_{10}tN^i_0A_i + \mathcal{M}^i\log_{10}r \\
\log_{10}N^i_0 &= \log_{10}N_0 + \eta(T^i - \mathrm{15:00}_\mathrm{UTC})
\end{array},\right.
\end{equation}
where $t$ is the duration of an observation unit, $A_i$ is the meteor-collecting area of the $i$-th observation unit in units of $\mathrm{km^2}$, $\mathcal{M}^i$ is a limiting magnitude in the $i$-th observation unit, $\eta$ is a parameter to describe the diurnal variation \citep{hawkins_variation_1956,hawkins_influx_1958,fentzke_latitudinal_2009} in units of $\mathrm{dex\,hour^{-1}}$, and $T_i$ is the observation time of the $i$-th observation unit in hours in UTC. From Equation~(\ref{eq:lf}), expected magnitudes of detectable meteors in the $i$-th observation unit $\tilde{M}^i$ should follow
\begin{equation}
\label{eq:magexp}
\mathcal{M}^i - \tilde{M}^i \sim \mathrm{exponential}(\beta),
\end{equation}
where $\beta$ is a shape parameter of the exponential distribution, given by $\beta = \log{r}$.
\begin{figure*}[tp]
\centering
\plotone{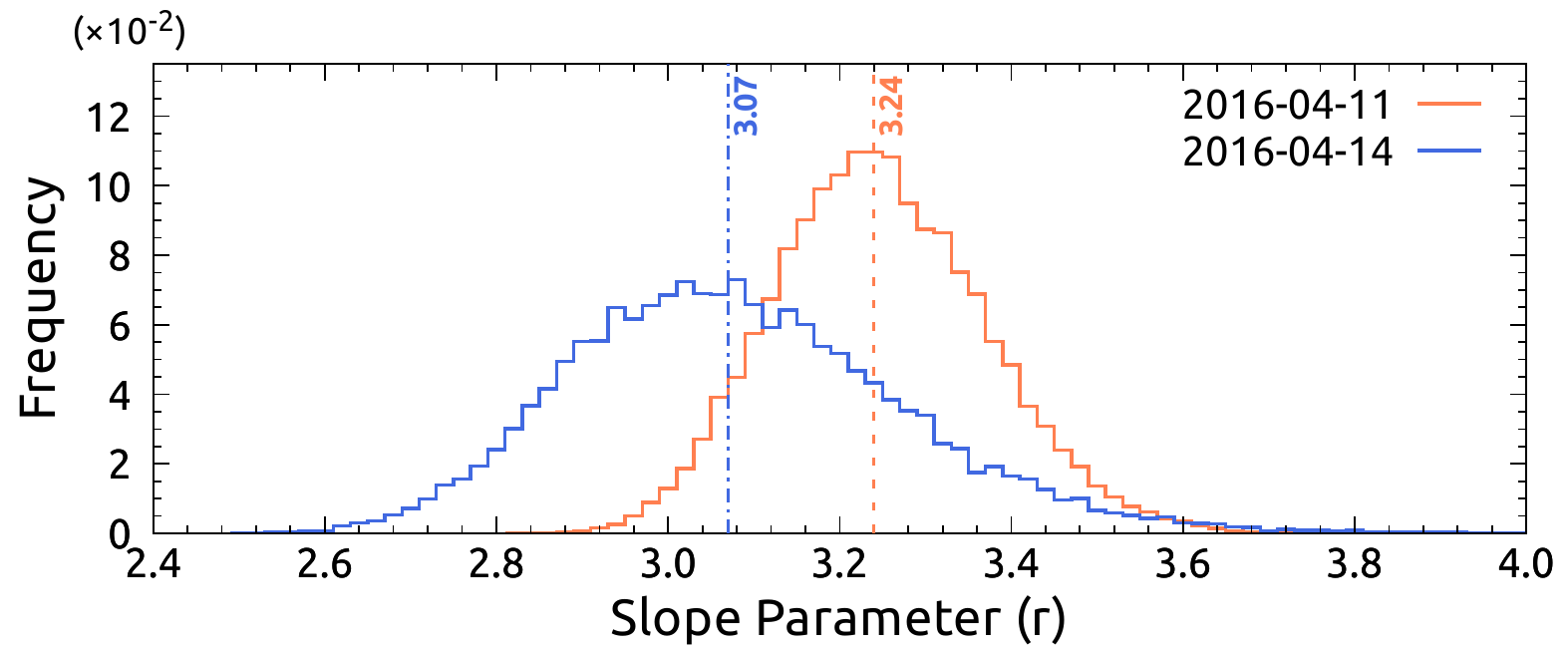}
\plotone{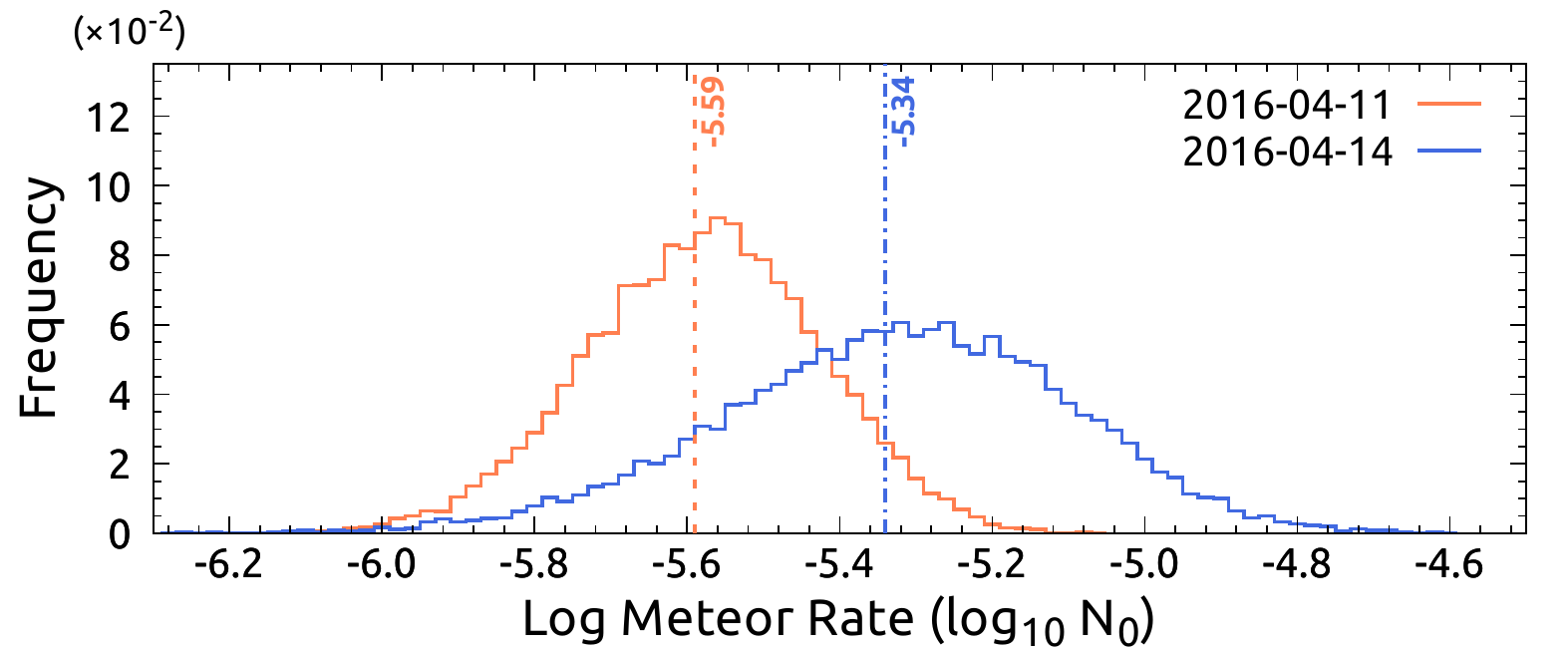}
\plotone{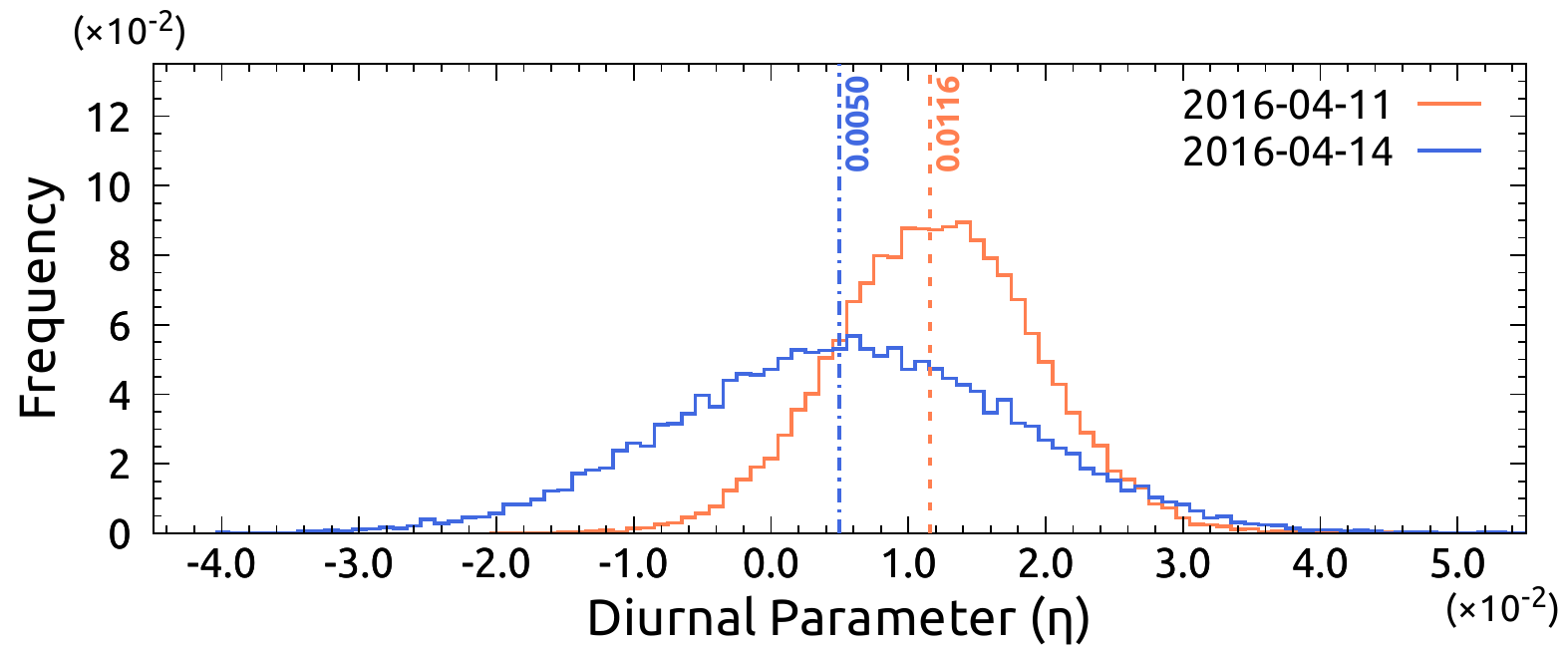}
\caption{Probability density funciton of the fitting parameters, $r$, $\log_{10}N_0$, and $\eta$, are shown in the top, middle, and bottom panels, respectively. The vertical lines with numbers indicate the posterior mean values.}
\label{fig:modelfit}
\end{figure*}
\begin{table*}
\centering
\caption{Statistical Inference with the Simple Model}
\label{tab:modelfit}
\begin{tabular}{cccccc}
\hline\hline
& Parameter & Mean & 2.5\% & 50\% & 97.5\% \\ \hline
2016-04-11
& $r$            &$\phm3.24$&$\phm3.01$&$\phm3.23$ &$\phm3.50$ \\
& $\log_{10}N_0$ &   $-5.59$&   $-5.91$&   $-5.59$ &   $-5.30$ \\
& $10^2\eta$     &$\phm1.16$&   $-0.30$&$\phm1.16$ &$\phm2.62$ \\
\hline
2016-04-14
& $r$            &$\phm3.08$&$\phm2.74$&$\phm3.06$&$\phm3.50$ \\
& $\log_{10}N_0$ &   $-5.34$&   $-5.82$&   $-5.33$&   $-4.93$ \\
& $10^2\eta$     &$\phm0.50$&   $-1.92$&$\phm0.50$&$\phm2.91$ \\
\hline
\end{tabular}
\end{table*}
The parameters are optimized to match $n^i \sim \tilde{n}^i$ and $M^i \sim \tilde{M}^i$. The optimization is carried out with \texttt{Stan} \citep{carpenter_stan:_2016}, which is a software for the Bayesian statistical inference with MCMC sampling \citep{carpenter_stan:_2016,homan_no-u-turn_2014}. The posterior probability distributions are shown in Figure~\ref{fig:modelfit}. The posterior mean values and the 95\% confidence intervals of the parameters are listed in Table~\ref{tab:modelfit}. No significant differences are detected between the results on April 11 and 14. The results suggest that the slope parameter is ${\sim}3.1\pm{0.4}$ and the meteor rate is about $-5.5{\pm}0.5$.
The data on April 11 marginally suggest a positive $\eta$ value, while the increase in the meteor rate is not confirmed on April 14. The present result, $\eta$ of ${\sim}1{\times}10^{-2}\,\mathrm{dex\,hours^{-1}}$, corresponds the increase by 30\% in 12 hours. The increase was much smaller than reported in previous studies \citep{murakami_annual_1955,kero_2009-2010_2012}. This could be in part attributable to our assumption in calculating the EMCA: the radiant distribution of meteors is uniform. Further observations are required to confirm the non-ditection of the diurnal variation with \tomoepm.
\begin{figure}[tp]
\centering
\plotone{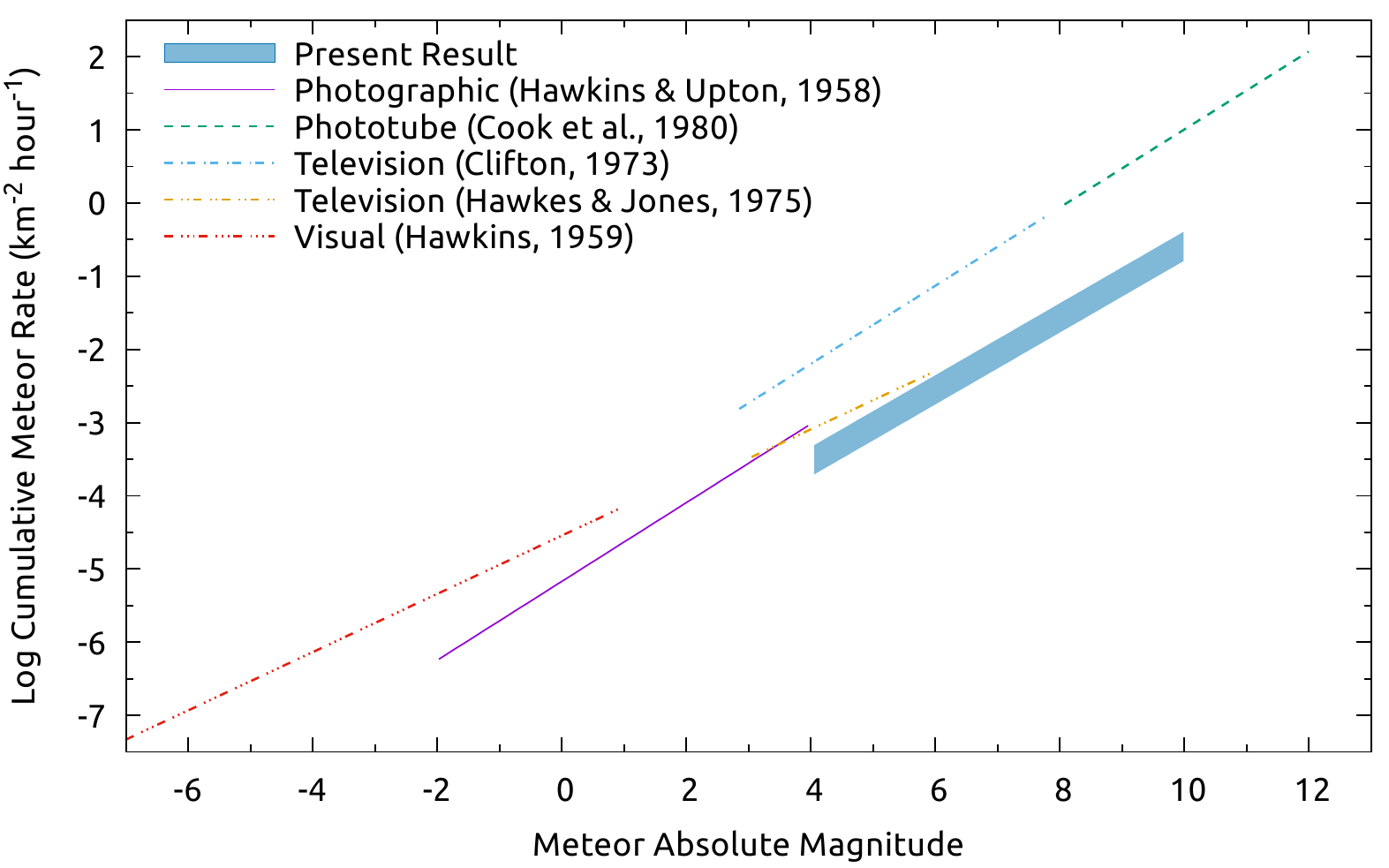}
\caption{Comparison of luminosity functions obtained in different observations. The present result is indicated by the blue region.}
\label{fig:luminosity_functions}
\end{figure}
We compare the present result with luminosity functions in literature in Figure~\ref{fig:luminosity_functions}. The purple solid line indicates the luminosity function derived by \citet{hawkins_influx_1958} based on surveys with Super-Schmidt cameras. \citet{hawkins_influx_1958} suggested that the $r=3.4{\pm}0.2$ and $\log_{10}N_0 = -5.2$ using meteors brighter than about $4\,\mathrm{mag}$. The green dashed line indicates the luminosity function provided by \citet{cook_flux_1980} using a 10-m reflector and phototubes. The slope parameter they derived was about $3.41$. The blue doted-dashed line and orange double-dotted-dashed line respectively show the luminosity functions in \citet{clifton_television_1973} and \citet{hawkes_inexpensive_1973}. They observed meteors using television systems. Since the brightness of meteors were not directly measured in \citet{clifton_television_1973}, we use the luminosity function re-calibrated by \citet{cook_flux_1980}. \citet{hawkes_television_1975} suggested that $r \sim 2.5$, while \citet{clifton_television_1973} suggested that $r \sim 3.4$. The red triple-dotted-dashed line shows the luminosity function of fireballs \citep{hawkins_relation_1959}, suggesting that $r \sim 2.5$. Here, we tentatively assume that the conversion factor from photographic to visual magnitude is ${+}1.0\,\mathrm{mag}$. The present result is shown by the blue region. The height of the region represents the uncertainties of the parameters.
The present observations constrain the luminosity function between about $+4$ to $+10\,\mathrm{mag}$, which is the deepest among the imaging observations. The slope parameter in the present result is roughly consistent with the other observations except for \citet{hawkins_relation_1959}. The meteor rate $N_0$ is by a factor of ${\sim}2.5$ lower than the value reported in \citet{hawkins_influx_1958}. This can be attributed to a seasonal variation in the sporadic meteor rate; The number of sporadic meteors in March and April is about a half of the annual average rate \citep{hawkins_variation_1956,kero_2009-2010_2012,murakami_annual_1955}. The meteor rate of the present result is by a factor of ${\sim}30$ lower than that of \citet{cook_flux_1980}. This difference could be in part attributable to the difference in the assumed EMCAs. \citet{cook_flux_1980} assumed that the EMCA was $3\,\mathrm{km^2}$, which was basically derived with the thin shell approximation and could be underestimated. Generally, the present result is consistent with the luminosity functions in literature. \citet{cook_flux_1980} suggested that the luminosity function of meteors was well approximated by a single slope power-law function from $-2.4$ to $+12\,\mathrm{mag}$. The present result is in line with the Cook's suggestion.
\subsection{System efficiency comparison between \tomoepm and \tomoe}
\begin{figure*}
\plotone{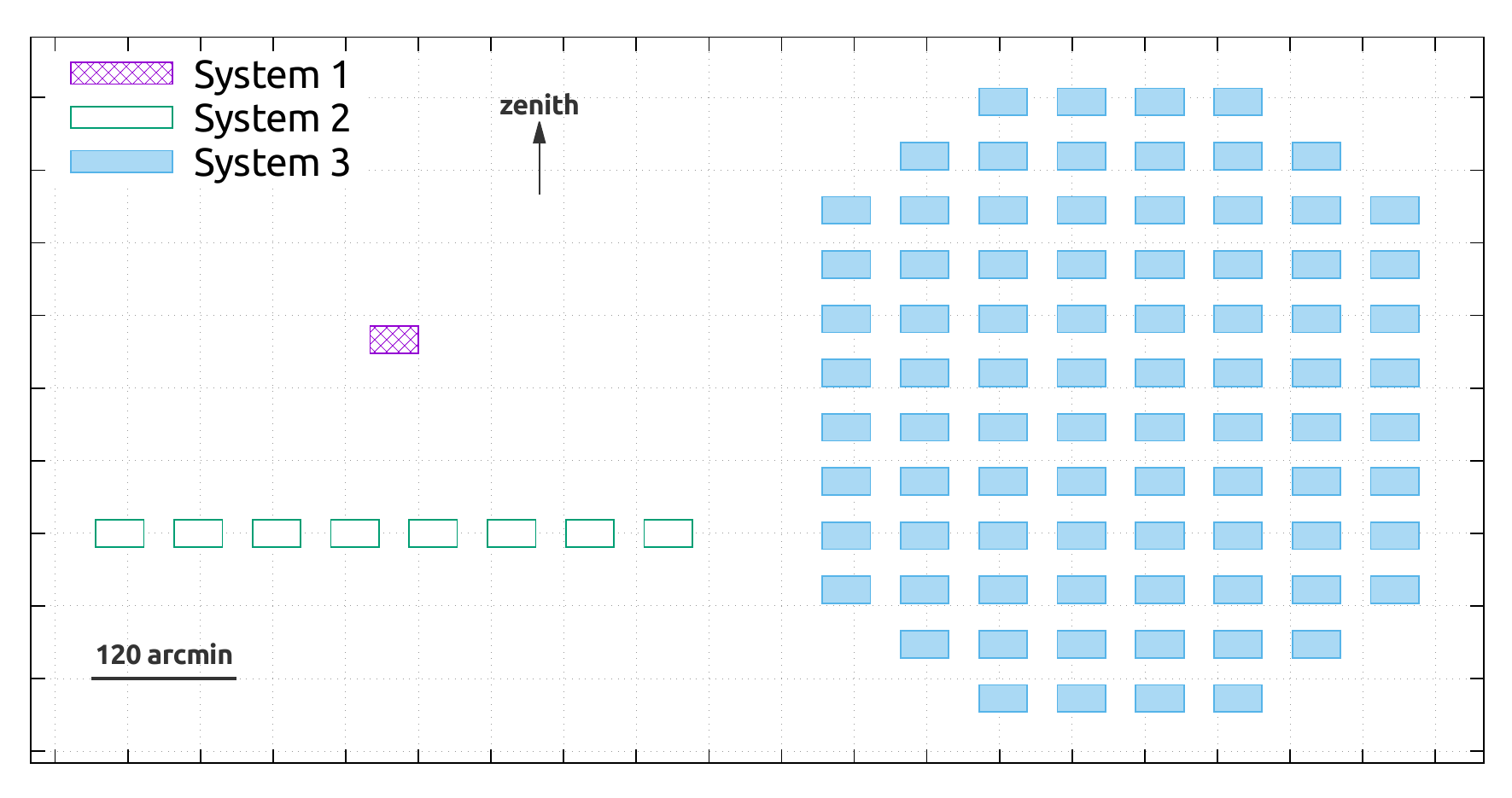}
\caption{\revise{20180710}{Fields-of-view} of the three systems. The violet cross-hatched region indicates the field-of-view of System 1. The green empty rectangles indicate the field-of-view of System 2. The blue filled rectangles indicate the field-of-view of System 3.}
\label{fig:sys}
\end{figure*}

\begin{figure*}
\plotone{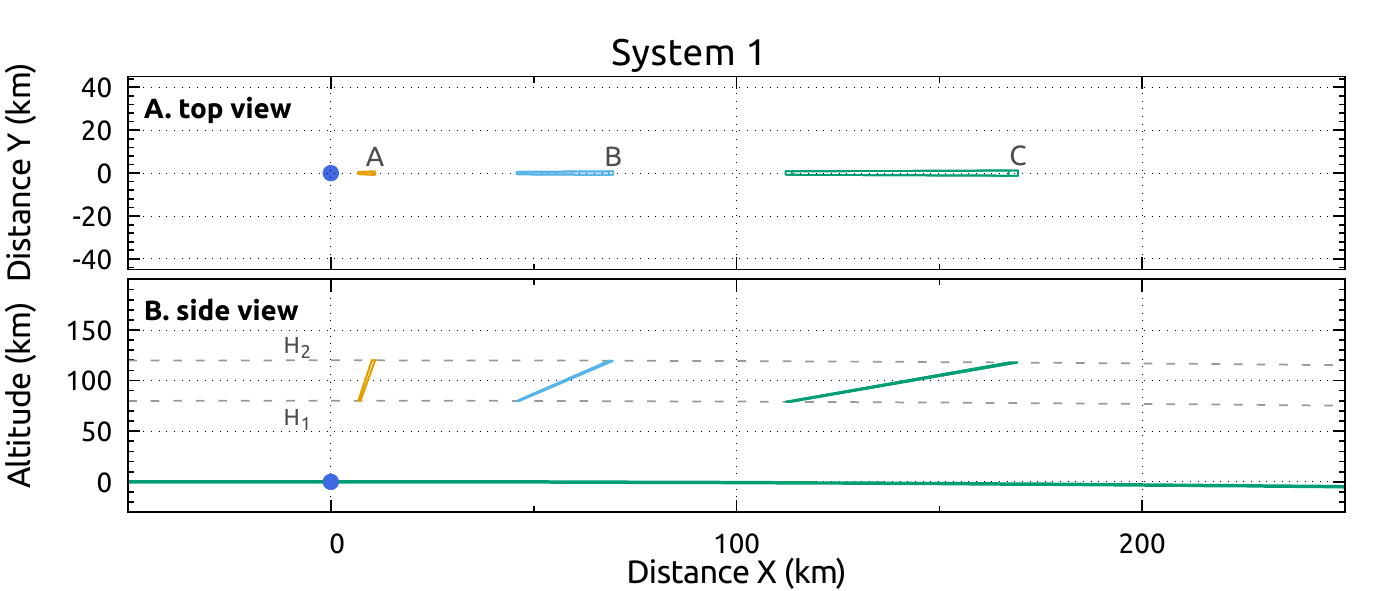}
\plotone{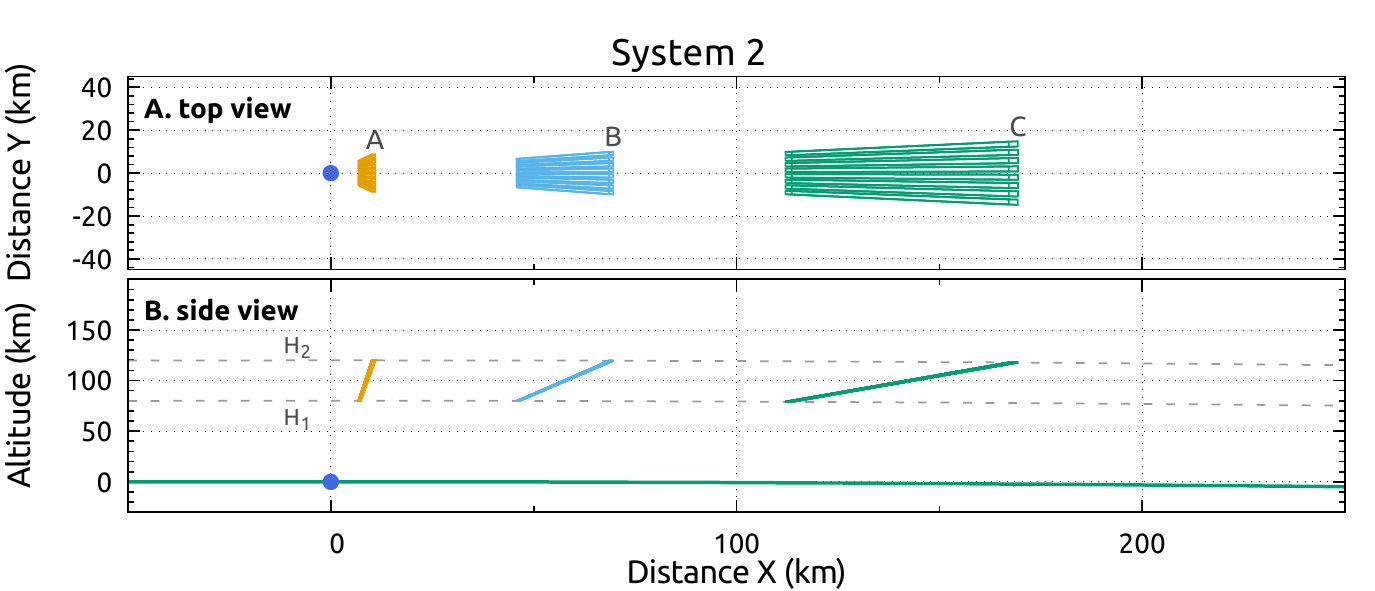}
\plotone{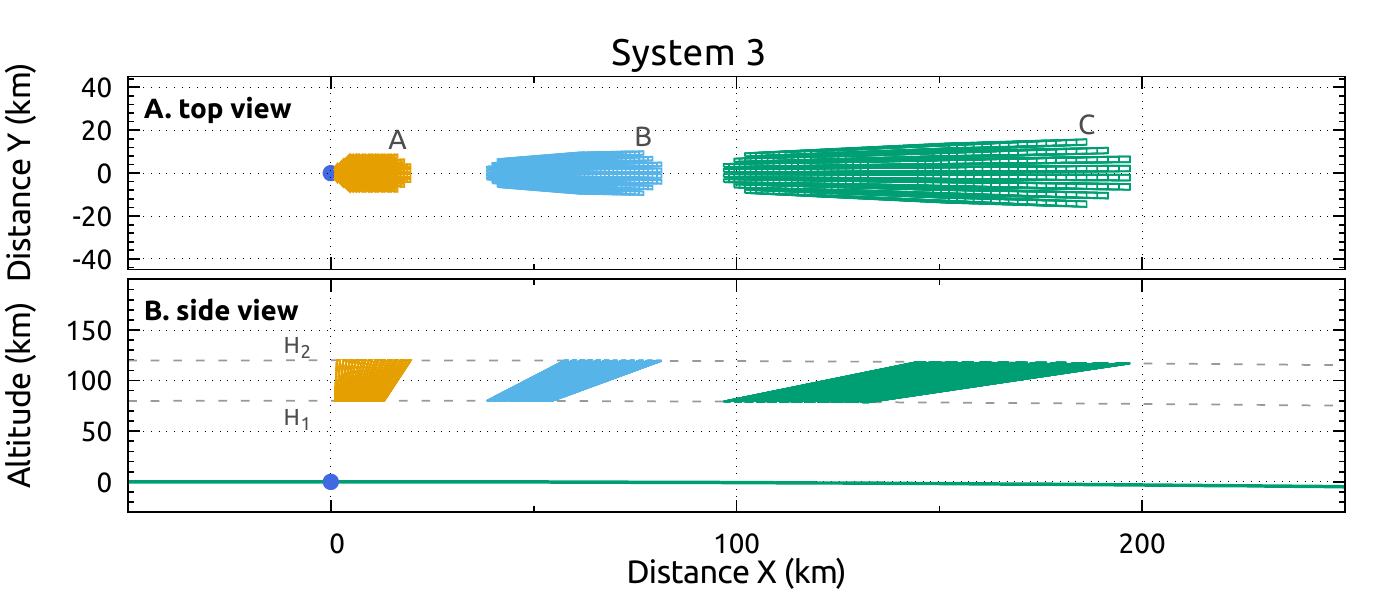}
\caption{Schematic view of meteor-collecting volumes (MCV). The top, middle, and bottom plots respectively show the MCVs for Systems 1, 2, and 3. The labels ``A'', ``B'', and ``C'' indicate the MCVs for elevation angles of $85\arcdeg$, $60\arcdeg$, and $35\arcdeg$, respectively. The blue dots indicate the location of the observing system. The green solid curve in Panels B indicates the sea level. The gray dashed lines in Panels B respectively show the top and bottom surface of the meteor sphere.}
\label{fig:mcv_sys1}
\end{figure*}

\begin{figure*}
\plotone{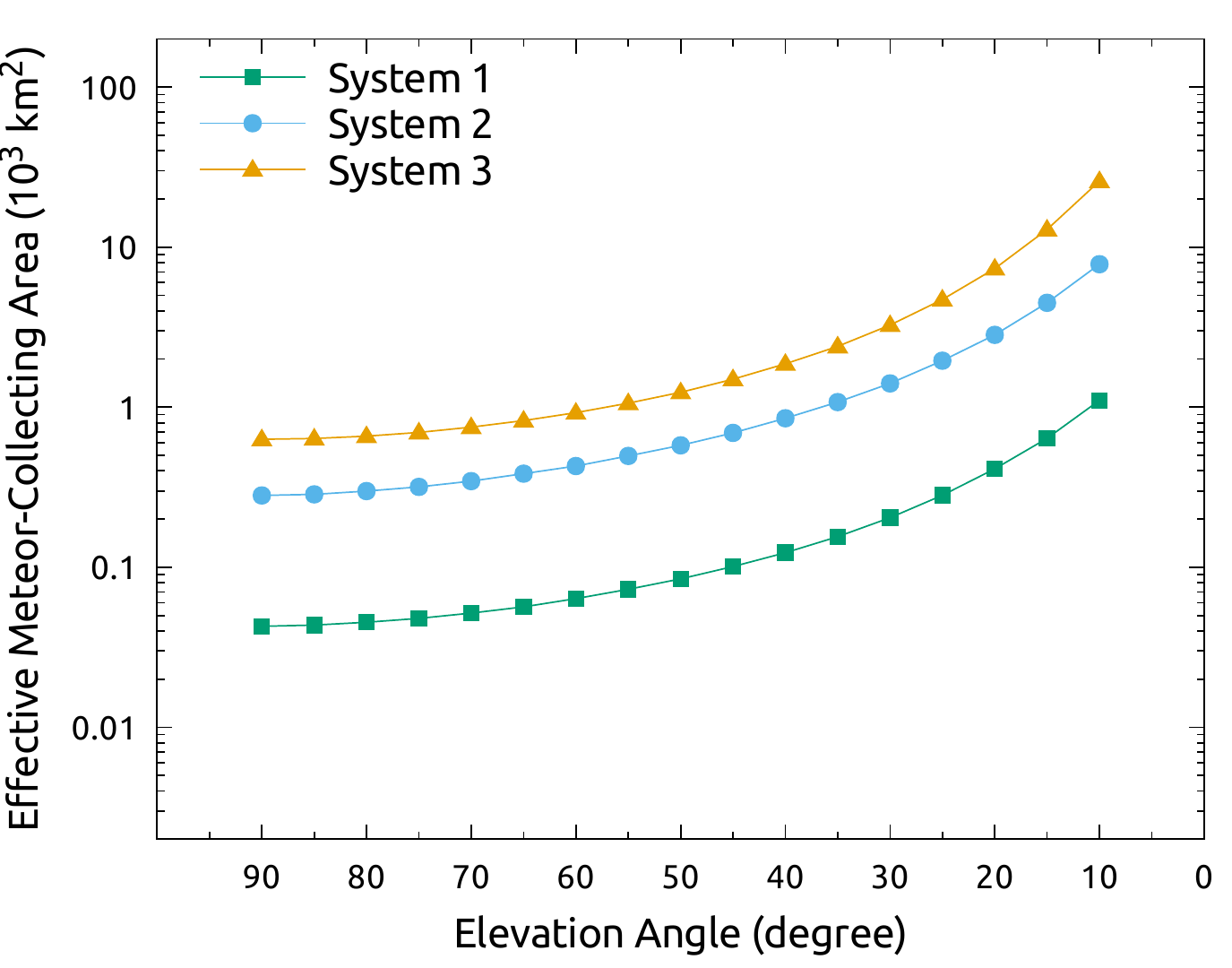}
\caption{Effective meteor-collecting areas (EMCAs) in units of $\mathrm{km^2}$ against an elevation angle $\alpha$. The error bars are as small as the line width.}
\label{fig:emca}
\end{figure*}

The operation of \tomoepm was completed in 2016. Observations with \tomoe started in February, 2018. Here, we compare the EMCAs between \tomoepm and \tomoe, to evaluate the efficiency in the meteor observations with \tomoe.
The EMCAs are calculated out for three systems. The shape of their \revise{20180710}{fields-of-view} are illustrated in Figure~\ref{fig:sys}. System 1 has a single field-of-view with dimensions of $39\arcmin.7{\times}22\arcmin.4$, corresponding to that of one CMOS sensor. System 2 emulates \tomoepm, a wide-field camera equipped with 8 CMOS image sensors mounted on 105\,cm Kiso Schmidt telescope \citep{sako_development_2016}. System 2 has eight image sensors, each of which has a field-of-view with dimensions of $39\arcmin.7 {\times} 22\arcmin.4$. System 3 is a counterpart of \tomoe. The field-of-view of System 3 is composed of 84 segments. The size of each segment is the same in Systems 1 and 2.
The MCVs with an elevation angle of $35\arcdeg$, $60\arcdeg$, and $85\arcdeg$ are illustrated in Figure~\ref{fig:mcv_sys1}. The position angles of the field-of-view are fixed. Both the size of the MCV and the distance to the MCV increase with decreasing elevation angle.
The EMCAs for the three systems are calculated for elevation angles from $10\arcdeg$ to $90\arcdeg$ at $5\arcdeg$ intervals. Figure~\ref{fig:emca} shows the dependence of the EMCA on the elevation angle $\alpha$. The EMCAs at zenith are about $43$, $281$, and $630\,\mathrm{km^2}$ for System 1, 2, and 3, respectively. At $\alpha = 10\arcdeg$, the EMCAs increase to about $1.1{\times}10^3$, $7.8{\times}10^3$, and $25.8{\times}10^3\,\mathrm{km^2}$ for System 1, 2, and 3, respectively. The EMCAs for Systems 1, 2, and 3 similarly increase with decreasing elevation angle. The dependence of the EMCA on $\alpha$ is well approximated by $\sin^{-2}\alpha$. \revise{20180710}{Note that the EMCAs, which do not include the effect of the distance to meteors, do not reflect the expected number of meteors detected. Refer to \ref{sec:rmca} for detailed discussion.}
The EMCAs for Systems 2 is about 7 times larger than that of System 1. This increase is almost proportional to the number of the image sensors. The efficiency of System 3 is about \revise{20180710}{two times} larger than that of System 2. Since the detectors of \tomoe are tiled in the image circle of the telescope, a large fraction of meteors will be detected first in peripheral detectors; the detectors around the center of the field-of-view have little contribution to the meteor-collecting efficiency.
\begin{table*}
\centering
\caption{EMCA for Square \revise{20180710}{Fields-of-View}}
\label{tab:emca_simple}
\begin{tabular}{p{9.0em}cccccc}
\hline\hline
& \multicolumn{1}{c}{$10\arcmin{\times}10\arcmin$}
& \multicolumn{1}{c}{$20\arcmin{\times}20\arcmin$}
& \multicolumn{1}{c}{$30\arcmin{\times}30\arcmin$}
& \multicolumn{1}{c}{$40\arcmin{\times}40\arcmin$}
& \multicolumn{1}{c}{$50\arcmin{\times}50\arcmin$}
& \multicolumn{1}{c}{$60\arcmin{\times}60\arcmin$}
\\\hline
Side Length (\arcmin)
& $10$
& $20$
& $30$
& $40$
& $50$
& $60$
\\
FOV area ($\mathrm{deg}^2$)
& $0.03$
& $0.11$
& $0.25$
& $0.44$
& $0.69$
& $1.00$
\\
EMCA ($\mathrm{km^2}$)
& $14.6$
& $27.2$
& $41.4$
& $55.1$
& $69.2$
& $83.2$
\\
Projected Area\tablenotemark{a} ($\mathrm{km^2}$)
& $0.09$
& $0.34$
& $0.76$
& $1.35$
& $2.12$
& $3.05$
\\
Event Rate\tablenotemark{b} ($\mathrm{hour^{-1}}$)
& $22$
& $45$
& $68$
& $90$
& $114$
& $136$
\\
\hline\hline
\end{tabular}
\tablenotetext{a}{The area of the field-of-view projected on the sphere at an altitude of $100\,\mathrm{km}$.}
\tablenotetext{b}{Estimated meteor event rates for meteors brighter than $10$ magnitude.}
\end{table*}
To investigate the dependence of the EMCA on the field-of-view, we calculate the EMCA for simple systems which have a single square field-of-view pointing at the zenith. The \revise{20180710}{side lengths} of the \revise{20180710}{fields-of-view} are $10\arcmin$, $20\arcmin$, $30\arcmin$, $40\arcmin$, $50\arcmin$, and $60\arcmin$. Calculated EMCAs are listed in Table~\ref{tab:emca_simple} as well as the areas of the \revise{20180710}{fields-of-view} projected on the surface at an altitude of $100\,\mathrm{km}$ above the sea level (\textit{hereafter}, referred as to projected \revise{20180710}{fields-of-view}).
The EMCA is larger than the area of the projected field-of-view. In the case of the $60\arcmin{\times}60\arcmin$ system, the EMCA is larger by a factor of about 30. This simply illustrates the thin shell approximation is not appropriate for systems with small \revise{20180710}{fields-of-view}.
The EMCAs are almost proportional to the \revise{20180710}{side lengths} of the \revise{20180710}{fields-of-view}, rather than the areas of the \revise{20180710}{fields-of-view}. \revise{20180710}{The meteor-collecting efficiency, in general, depends on the surface area of the MCV. For a narrow field-of-view, the surface area of the MCV is approximately proportional to the side length of the field-of-view. This simply explains the dependence of the EMCA on the side length; The EMCA of narrower field-of-view systems does not decrease as with the area of the field-of-view. This is generally consistent with the fact that the EMCA of System 3 is about two times larger than that of System 2; The summation of the most outer circumference sides of System 2 is about $11.3\arcdeg$, while that of System 3 is about $19.5\arcdeg$.}
The expected number of detected meteors per hour is estimated by $A N_0 r^{\mathcal{M}}$, where $\mathcal{M}$ is a limiting magnitude, $r$ is a meteor index, $A$ is an EMCA, and $N_0$ is a rate of meteors brighter than zeroth magnitude. Here, we adopt $r=3.4$ and $\log_{10}N_0 = -5.1$ \citep{hawkins_influx_1958}. Table~\ref{tab:emca_simple} lists the event rate calculated for $\mathcal{M} = 10\,\mathrm{mag}$ in optical wavelengths. A system with a $1\arcdeg{\times}1\arcdeg$ field-of-view and a limiting magnitude of ${\sim}10$ will detect more than 100 meteors per hour. This result encourages meteor survey observations with wide-field cameras mounted on large aperture telescopes (e.g. Subaru/HSC \citep{miyazaki_hyper_2012}).
\tomoe is expected to detect more than about 2,000 faint meteors a night. Since meteors will run across multiple detectors of \tomoe, faint meteors are more robustly detected with \tomoe than \tomoepm. \tomoe will be an ideal instrument to investigate variations or structures in the luminosity function of meteors brighter than about ${+}10\,\mathrm{mag}$.
\section{Conclusion}
\label{sec:conclusion}
\tomoe is a mosaic CMOS camera developed in Kiso Observatory, the University of Tokyo. \tomoe, equipped with 84 CMOS sensors, will continuously obtain images of $20\,\mathrm{deg^2}$ at $2\,\mathrm{Hz}$. In this sense, \tomoe will be the world largest video camera for astronomy. As a pilot project, we developed a prototype model of \tomoe, \tomoepm, which has 8 CMOS sensors. \tomoepm can monitor a sky of about $2\,\mathrm{deg^2}$ at $2\,\mathrm{Hz}$. We carried out imaging observations of faint meteors with \tomoepm on April 11 and 14, 2016. We set the observation field within the Earth's shadow to eliminate contamination from debris and satellites in the low Earth orbit. The total observing time was about 10 hours, and 2,220 meteor events were detected in total.
\revise{20180710}{Our observations provide a new measurement of the meteor luminosity function.} The video rate magnitudes of the meteors we detected are typically from $+4$ to $+12\,\mathrm{mag}$. The corresponding mass of the meteors ranges from $8.3{\times}10^{-2}$ to $3.3{\times}10^{-4}\,\mathrm{g}$. \revise{20180710}{The present results are consistent with a single power-law luminosity function.} A statistical model suggest that the slope parameter $r = 3.1{\pm}0.4$ and the meteor rate $\log_{10}N_0 = -5.5{\pm}0.5$. The diurnal variation in the sporadic meteor rate is marginally confirmed only on April 11. \revise{20180710}{The slope is roughly consistent with those of luminosity functions in literature. The meteor rate is lower than those in literature. The discrepancy is, in part, attributed to the seasonal variation and different assumptions on the EMCAs.}
The operation of \tomoepm was completed in 2016. Observations with \tomoe started in February, 2018. We compare the effective meteor-collecting areas between \tomoepm and \tomoe, suggesting that \tomoe is twice as efficient as \tomoepm in detecting meteors. More than 2,000 meteors will be detected every night with \tomoe. Kiso Observatory has \revise{20180710}{the capability to make it a leading place} in time-domain astronomy including observations of faint meteors.
\section*{Acknowledgments}
This research is supported in part by JSPS Grants-in-Aid for Scientific Research (KAKENHI): Grant Numbers JP16H02158, JP16H06341, JP18H01261, and JP18K13599. This research is in part supported by JSPS Grant-in-Aid for Scientific Research on Innovative Areas: Grant Number JP2905. This is also supported in part by PRESTO, Japan Science and Technology Agency (JST). This work is also supported by the Optical and Near-infrared Astronomy Inter-University Cooperation Program by the MEXT of Japan. This work is achieved using the grant of Joint Development Research by the Research Coordination Committee, National Astronomical Observatory of Japan (NAOJ). The fabrication of Tomo-e PM is conducted in collaboration with the Advanced Technology Center of NAOJ. This research is supported by Canon Inc. for developing the full-HD high-sensitivity CMOS sensors.

\clearpage
\appendix
\section{Validation of the EMCA Calculation}
\label{sec:emca_validation}
\subsection{Travel Distances}
\label{sec:travel_distances}
\begin{figure*}[p]
\plotone{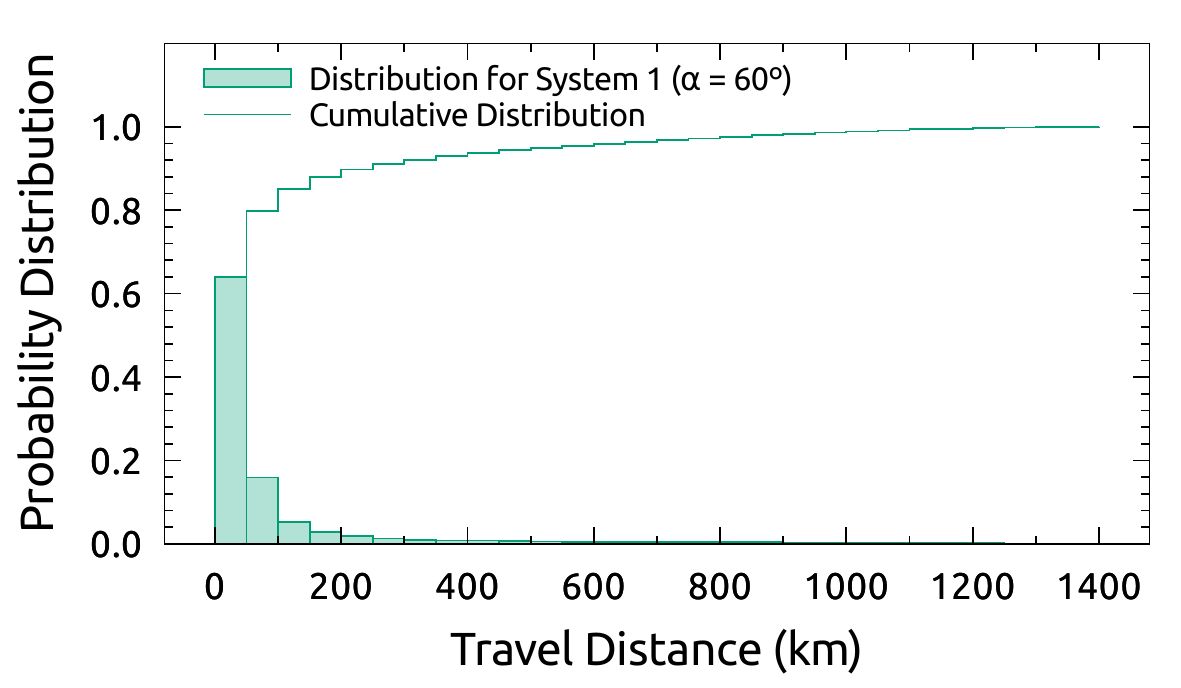}
\caption{Distribution of the travel distances of observable meteors. The filled bars show the histogram the travel distances for System 1 with an elevation angle $\alpha = 60\arcdeg$. The cumulative distribution is described by the solid step line.}
\label{fig:travel}
\end{figure*}

In the present formalism, we assume that an incident meteor never fades away inside the meteor shell. This assumption may bring an overestimate of a rate of meteors with a long travel distance, and an EMCA could be overestimated accordingly. Figure \ref{fig:travel} shows the distribution of the travel distances of the meteors for System 1 with $\alpha = 60\arcdeg$. The distribution neither depends on the systems nor elevation angles. The fraction of the meteors with a travel distance longer than $400\,\mathrm{km}$ was $\lesssim 10\%$. Thus, we concluded that the assumption had virtually no effect on the present results.
\subsection{Position Angles}
\label{sec:position_angles}
\begin{figure*}[p]
\plotone{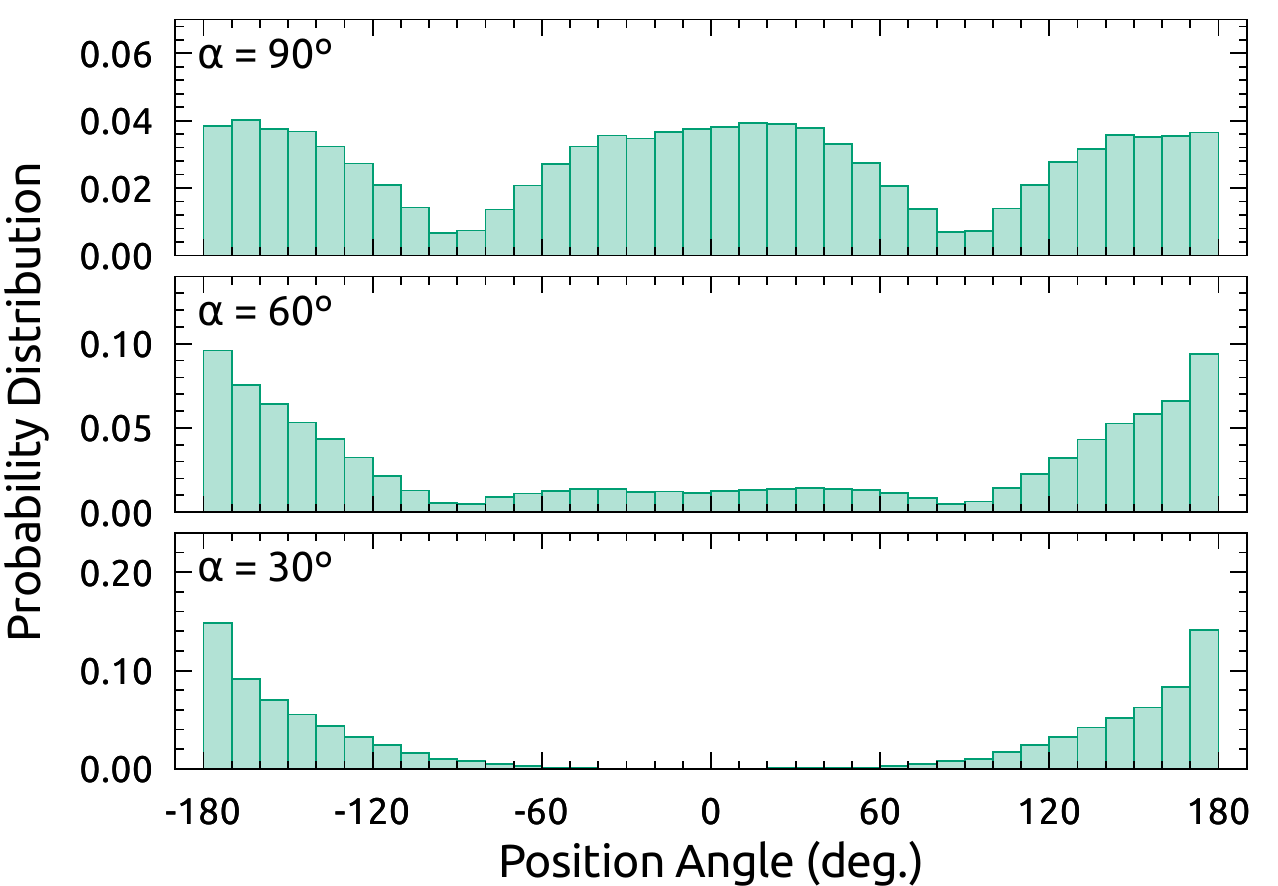}
\caption{Distributions of the position angles of mock meteors. A position angle of zero corresponds to a meteor moving toward the zenith, or upward in the field-of-view.}
\label{fig:pa}
\end{figure*}

Artificial patterns are found in the distribution of position angles on the focal plane. Position angles (PA) of meteors for System 2 are summarized in Figure~\ref{fig:pa}. Dips around $\mathrm{PA} = -90\arcdeg$ and $90\arcdeg$ in the top panel of Figure~\ref{fig:pa} are attributed to the alignment of the detectors. A fraction of meteors with $\mathrm{PA} \sim {{\pm}180\arcdeg}$ increased with decreasing elevation angle. Thus, a large fraction of meteors run from the zenith to the nadir in a field-of-view. The observations on April 11 and 14 were carried out with elevation lower than 38\arcdeg. Similar distributions of the PAs were confirmed in the observations. Our calculation is qualitatively consistent with the observations. The degree of the concentration depends on the distribution of the incident angle $i$. Since it was difficult to constrain the distribution of $i$ from the observations, we did not discuss the PA distribution of the meteors in the paper.
\subsection{Depth of the Meteor Shell}
\label{sec:depth_meteor_shell}
We assume that the top and bottom boundaries of the meteor shell were at $120$ and $80\,\mathrm{km}$, respectively. They are chosen as nominal values \citep{jenniskens_meteor_2006}. An appearing and disappearing altitude of a meteor, however, depend on the size or composition of the meteoroid. We calculate the EMCAs for $H_1 = 60\,\mathrm{km}$ and $H_2 = 140\,\mathrm{km}$, the results are almost the same. Thus, we conclude that the choice of $H_1$ and $H_2$ has little impact on the present results.
\subsection{Reduced Efficient Meteor-Collecting Area}
\label{sec:rmca}
\begin{figure*}[p]
\plotone{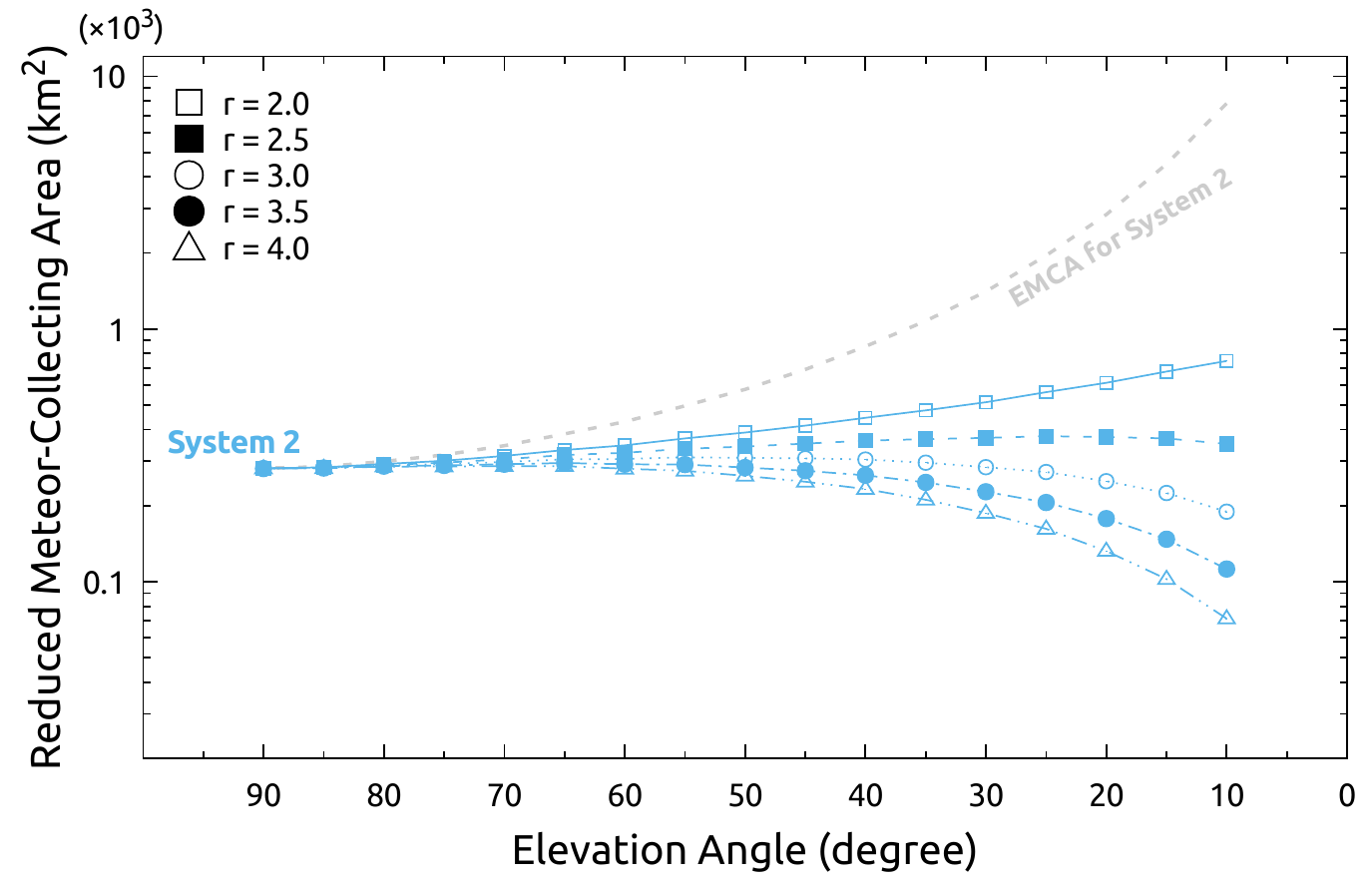}
\caption{Reduced effective meteor-collecting areas (RMCAs) in units of $\mathrm{km^2}$ against an elevation angle $\alpha$. The symbols indicate an assumed meteor index value $r$. The gray dashed line indicates the EMCA for System 3 as a reference.}
\label{fig:rmca}
\end{figure*}

\begin{figure*}[p]
\plotone{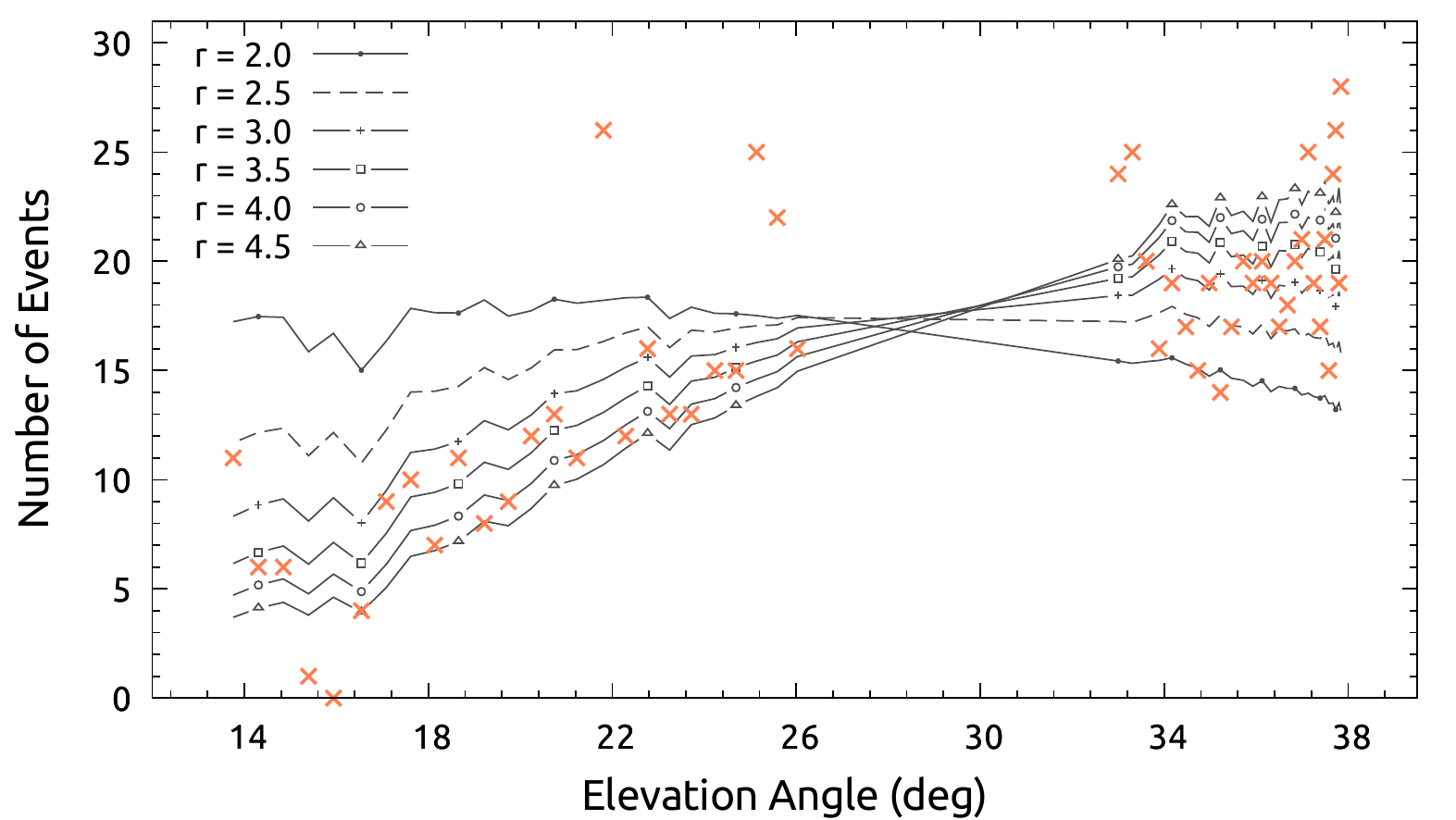}
\caption{Deriving the meteor index $r$ by counting the number of meteors. The orange crosses shows the number of the meteors in 3\,minute against the elevation angle in observation. The lines show the expected number of detections for different $r$ values.}
\label{fig:index}
\end{figure*}

The EMCA defined in Equation~(\ref{eq:mca}) does not take into account the apparent brightness of incident meteors and atmospheric extinction. A practical meteor-collecting area is given by reducing $A$ as
\begin{equation}
\label{eq:red}
A^\mathrm{red} =
\frac{A_0}{N_0}\left(\frac{100\,\mathrm{km}}{H_2}\right)^2
\sum^N_i r^{-5\log_{10}\frac{d_i}{100\,\mathrm{km}}-\epsilon_i},
\end{equation}
where $r$ is the population index, $d_i$ is the distance to the $i$-th meteor, and $\epsilon_i$ is the amount of extinction of the $i$-th observable meteor. The reduced effective meteor-collecting area (\textit{hereafter}, referred as to RMCA) corresponds to the area defined in Equation~(1) of \citet{koschack_determination_1990}. For the sake of simplicity, we adopt $\epsilon_i = 0$ in the rest of the paper.
We calculate the RMCAs for Systems 2. The position angle $\phi$ is fixed at zero. Figure~\ref{fig:rmca} shows the RMCA for $r=2.0$, $2.5$, $3.0$, $3.5$, and $4.0$ against an elevation angle. While the RMCA is almost the same as the EMCA when the elevation angle $\alpha$ is larger than about $80\arcdeg$, the differences between the EMCA and RMCA become significant at low elevation angles.
The differences increase as $r$ becomes large. \citet{hawkins_influx_1958} and \citet{cook_flux_1980} independently obtained $r \simeq 3.4$ for sporadic meteors with $M_\mathrm{pg} < {+}5$ and ${+}7 < M_\mathrm{pg} < {+}12$, respectively. Thus, we safely assume $r > 3.0$ for visible sporadic meteors. In such cases, the RMCA is largest at the zenith. This does not depend on systems. In observations of sporadic meteors, a monitoring observation close to the zenith is favored. When $r$ is larger than about $3.0$, the RMCA monotonically decreased with decreasing elevation angle in contrast to the EMCA. The RMCA is well-approximated by
\begin{equation}
\label{eq:approxAred}
\tilde{A}^\mathrm{red} =
Ar^{-5\log_{10}\frac{\langle d \rangle}{100\,\mathrm{km}}},
\end{equation}
where $\langle d \rangle$ is the averaged distance to the MCVs. Since the field-of-view is sufficiently narrow, the distances to the meteors are approximately identical. This justifies Equations (\ref{eq:expnum}) and (\ref{eq:approxAred}).
Equation~(\ref{eq:red}) indicates that an expected number of meteors can be given by $\tilde{n} = A^\mathrm{red}N_0r^\mathcal{M}$, which is a function of the meteor index $r$, the elevation angle $\alpha$, and the limiting magnitude $\mathcal{M}$. Figure~\ref{fig:rmca} illustrates the dependence of $A^\mathrm{red}$ on the meteor index $r$; the RMCA ratio at $\alpha = 30\arcdeg$ to $\alpha = 90\arcdeg$ decreases from 1.74 to 0.63 when the meteor index $r$ increases from 2.0 to 4.0. Thus, comparing the number of meteors detected in different elevation angles provides an estimate of the meteor index $r$ without measuring the brightness of the meteors. Figure~\ref{fig:index} shows an application of this method. The orange crosses show the numbers of meteor events in 3\,minute on April 11, 2016\footnote{The data on April 14, 2016 are not shown, since large part of the data with $\alpha > 30\arcdeg$ were obtained in bad conditions.}. The number of detections decreases with decreasing elevation angle. The lines indicate the expected number of meteor detections for different $r$ values. The limiting magnitudes are estimated from neighboring stars. The observations are well explained by the expectations with $r=3.0$--$4.0$, consistent with previous studies \citep{cook_flux_1980,hawkes_television_1975,hawkins_influx_1958}. This confirms that the present results are self-consistent.
\end{document}